\documentclass[aps,prd,reprint,floatfix,superscriptaddress,lengthcheck,sort&compress,letterpaper,nofootinbib]{revtex4-2}

\usepackage{amssymb}
\usepackage[intlimits]{amsmath}
\usepackage{amstext} % for \text macro
\usepackage{amsfonts}
\usepackage{dsfont}
\usepackage{float}
\usepackage{slashed}
\usepackage{subfigure}
\usepackage{natbib}
\usepackage{multirow}
\usepackage{verbatim}
\usepackage{nicefrac}
\usepackage{dcolumn}
\usepackage{units}
\usepackage{bm}
\usepackage{wasysym}
\usepackage{array}
\usepackage{graphicx}
\usepackage[usenames]{color}
\usepackage[colorlinks=true,linkcolor=blue,citecolor=blue,urlcolor=blue]{hyperref}
\usepackage{tabularx}
\usepackage{makecell}

\def\mG{\ensuremath{\mathcal{G}}}

\def\mN{\ensuremath{\mathcal{N}}}

\def\mR{\ensuremath{\mathcal{R}}}

\newcommand{\vect}[1]{{\mbox{\boldmath $#1$}}}

      \definecolor{violet}{RGB}{111,0,255}
      \definecolor{webgreen}{rgb}{0,0.75,0}
      \definecolor{webred}{rgb}{0.75,0,0}
      \definecolor{webblue}{rgb}{0,0,0.75}
      \definecolor{darkblue}{rgb}{0,0,0.6}
      \definecolor{darkgreen}{rgb}{0,0.5,0.5}
      \definecolor{darkpurple}{rgb}{0.5,0,0.5}
      \definecolor{darkorange}{rgb}{1,0.5,0}
      \definecolor{darkgrey}{rgb}{0.4,0.4,0.4}
      \definecolor{lgray}{rgb}{0.95,0.95,0.95}
      \definecolor{lgreen}{rgb}{0.95,1.00,0.90}
      \definecolor{lred}{rgb}{1.00,0.90,0.80}
      \definecolor{lblue}{rgb}{0.2,0.35,1.00}
      \definecolor{shadecolor}{rgb}{1.00,0.92,0.82}

\graphicspath{
    {./}
	{./figures/}
	{../figures/}
}

\begin{document}

\title{Going to the light front with contour deformations}

\author{Gernot Eichmann}
\email{gernot.eichmann@tecnico.ulisboa.pt}
\affiliation{LIP Lisboa, Av.~Prof.~Gama~Pinto 2, 1649-003 Lisboa, Portugal}
\affiliation{Departamento de F\'isica, Instituto Superior T\'ecnico, 1049-001 Lisboa, Portugal}

\author{Eduardo Ferreira}
\email{eduardo.b.ferreira@tecnico.ulisboa.pt}
\affiliation{LIP Lisboa, Av.~Prof.~Gama~Pinto 2, 1649-003 Lisboa, Portugal}
\affiliation{Departamento de F\'isica, Instituto Superior T\'ecnico, 1049-001 Lisboa, Portugal}

\author{Alfred Stadler}
\email{stadler@uevora.pt}
\affiliation{Departamento de F\'isica, Universidade de \'Evora, 7000-671 \'Evora, Portugal}
\affiliation{LIP Lisboa, Av.~Prof.~Gama~Pinto 2, 1649-003 Lisboa, Portugal}
\affiliation{Departamento de F\'isica, Instituto Superior T\'ecnico, 1049-001 Lisboa, Portugal}

\begin{abstract}
      We explore a new method to calculate the valence light-front wave function of a system of two interacting particles, which is based on
      contour deformations combined with analytic continuation methods to
      project the Bethe-Salpeter wave function onto the light front. In this proof-of-concept study, we
      solve the Bethe-Salpeter equation for a scalar model and find excellent agreement between the light-front wave functions obtained with contour deformations
      and those obtained with the Nakanishi method frequently employed in the
      literature. The contour-deformation method is also able to handle extensions of the scalar model that mimic certain features of QCD
      such as unequal masses and complex singularities.
      In principle the method is  suitable for computing parton distributions on the light front such as PDFs, TMDs and GPDs in the future.
\end{abstract}

\maketitle

\section{Introduction}

\begin{figure}[b]
                    \begin{center}
                    \includegraphics[width=0.74\columnwidth]{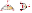}
                    \caption{Pictorial representation of the hadron-to-hadron correlator in Eq.~\eqref{2pt-correlator} (left) and the
                             Bethe-Salpeter wave function in Eq.~\eqref{bswf} (right).}\label{fig:correlators}
                    \end{center}
                    \vspace{-5mm}
            \end{figure}

   Understanding the quark-gluon structure of hadrons is
   a major goal in strong interaction studies.
   Ongoing and future experiments at the LHC, Jefferson Lab, RHIC, the Electron-Ion Collider, COMPASS/AMBER and other facilities aim to
   establish a three-dimensional spatial imaging of hadrons and
   measure structure observables such as
   the spin and orbital angular momentum distributions inside hadrons and their longitudinal and
   transverse momentum structure.
   These properties are encoded in
   PDFs (parton distribution functions), GPDs (generalized parton distributions)
   and TMDs (transverse momentum distributions), see e.g.~\cite{Diehl:2003ny,Belitsky:2005qn,Boer:2011fh,Lorce:2011dv,Accardi:2012qut} and references therein, whose  matrix elements
     \begin{equation}\label{2pt-correlator}
            \mG(z,P,\Delta) = \langle P_f | \,\mathsf{T}\,\Phi(z)\,\mathcal{O}\,\Phi(0)\,|P_i\rangle
       \end{equation}
   are illustrated in Fig.~\ref{fig:correlators}.
   We denoted the field operators generically by $\Phi(z)$, $\mathsf{T}$ denotes time ordering, $\mathcal{O}$ is some operator (usually containing a Wilson line),
   $\Delta$ is the momentum transfer, and
   $P_{f,i} = P \pm \Delta/2$ are the final and initial momenta of the hadron.

   A common feature of parton distributions is that they
   are defined  on the light front,
   i.e., the partons (quarks and gluons) inside the hadron are probed
   at a lightlike separation. In light-front coordinates this amounts to $z^+ = 0$, which
   by a Fourier transform
   translates to an integration over $q^-$ in momentum space, where $q$ is the relative momentum between the probed partons.
   In principle, the various quantities of interest can then be derived from Eq.~\eqref{2pt-correlator} as indicated in Table~\ref{tab:GPDs}:
   After taking the forward limit $\Delta = 0$, TMDs follow from an integration over $q^-$ and PDFs from another integration over
   the transverse momentum $\vect{q}_\perp$; for $\Delta \neq 0$ the same steps lead to generalized TMDs and GPDs~\cite{Lorce:2011dv,Accardi:2012qut}.

   In practice,
   taking Eq.~\eqref{2pt-correlator} to the light front is difficult in Euclidean formulations
   such as lattice QCD and continuum functional methods.
   To this end, lattice QCD has seen major recent progress in calculating quasi-PDFs, pseudo-PDFs and other quantities connected to Eq.~\eqref{2pt-correlator} from the path integral
   which allow one to reconstruct the desired parton distributions; see e.g.~\cite{Ji:2013dva,Ji:2014gla,Radyushkin:2016hsy,Radyushkin:2017cyf,Chen:2016utp,Orginos:2017kos,Lin:2017snn,Bali:2017gfr,Alexandrou:2018pbm,Ji:2020ect,Constantinou:2020hdm}.

   Functional methods, on the other hand, are usually formulated in momentum space.
   Here the matrix element~\eqref{2pt-correlator} needs to be constructed in a consistent manner from the elementary $n$-point correlation functions
   along the lines of Refs.~\cite{Tiburzi:2001je,Kvinikhidze:2004dy,Eichmann:2011ec,Nguyen:2011jy,Mezrag:2014jka,Mezrag:2016hnp,Bednar:2018mtf,Ding:2019lwe,Freese:2020mcx}.
   However, the integration over $q^-$ is not straightforward unless the analytic structure of the integrands is fully known, which is only the case in simple models
   where one may employ residue calculus, Feynman parametrizations or similar methods.

   In this respect, the Nakanishi integral
   representation
   has proven very efficient in recent years~\cite{Nakanishi:1963zz,Nakanishi:1969ph,Nakanishi:1988hp,Kusaka:1995za,Kusaka:1997xd,Sauli:2001we,Karmanov:2005nv,Sauli:2008bn,Carbonell:2010zw,Frederico:2011ws,Frederico:2013vga,Gutierrez:2016ixt,dePaula:2016oct,dePaula:2017ikc,AlvarengaNogueira:2019zcs}.
   Here the idea is to recast the hadronic amplitudes that appear in the integrands
   in terms of a weight function with a denominator that absorbs the analytic structure.
   There remain however questions regarding the formulation of generalized spectral representations for gauge theories,
   and the singularity structure of the remaining parts of the integrands (propagators, vertices, etc.) must still be known explicitly
   which poses practical limitations.

   The combination of such techniques, occasionally together with a reconstruction using moments, has found widespread recent applications in the calculation of parton distributions and related
   quantities~\cite{Miller:2009fc,Frederico:2010zh,Nguyen:2011jy,Bashir:2012fs,Chang:2013pq,Frederico:2013vga,Cloet:2013tta,Mezrag:2014jka,Chang:2014lva,Mezrag:2016hnp,Gutierrez:2016ixt,dePaula:2016oct,Chen:2016sno,dePaula:2017ikc,Mezrag:2017znp,Xu:2018eii,Bednar:2018mtf,Shi:2018zqd,AlvarengaNogueira:2019zcs,Freese:2019eww,Freese:2020mcx,Serna:2020txe,Ding:2019lwe,Ydrefors:2020duk,Shi:2020pqe,Zhang:2021mtn,Ydrefors:2021mky}.

            \begin{table}[t] \renewcommand{\arraystretch}{1.2}
            \begin{center}
            \begin{tabular}{r @{\quad} | @{\quad} c @{\quad\;} c @{\quad\;} c @{\quad} }
                                      & $\mG(q,P,\Delta=0)$   &  $\mG(q,P,\Delta)$  &  $\Psi(q,P)$  \\[1mm] \hline \rule{-0.8mm}{0.4cm}

            $\int dq^-$            &  TMD                   & GTMD      &   LFWF               \\
            $\int d^2\vect{q}_\perp \int dq^-$ &  PDF       & GPD       &  PDA

            \end{tabular}
            \end{center}
            %\vspace{-2mm}
            \caption{Light-front quantities following from the correlators~(\ref{2pt-correlator}--\ref{bswf})
                     after integrating over $q^-$ and $\vect{q}_\perp$.}
            \label{tab:GPDs}
            \end{table}

  \pagebreak

  The goal of the present work is to explore a new technique to compute correlation functions on the light front directly. It is
  based on contour deformations and analytic continuations, and in principle it does not rely on  explicit knowledge of the analytic structure of the integrands
  except for certain kinematical constraints.
  Instead of the hadron-to-hadron correlator~\eqref{2pt-correlator}, we consider the simpler case of the vacuum-to-hadron amplitude  shown in Fig.~\ref{fig:correlators},
  \begin{equation}\label{bswf}
     \Psi(z,P) = \langle 0 | \mathsf{T}\,\Phi(z)\,\Phi(0) | P \rangle \,.
  \end{equation}
  This is the generic form of a two-body Bethe-Salpeter wave function (BSWF) for a hadron carrying momentum $P$,
  which can be dynamically calculated from its Bethe-Salpeter equation (BSE).
  Taking this object onto the light front by integrating over $q^-$
  gives the valence light-front wave function (LFWF), cf.~Table~\ref{tab:GPDs}, which will be the central object of interest in this work to establish and test the method.
  In light-front quantum field theory, the LFWFs are the coefficients of a Fock expansion and thus acquire a probability interpretation~\cite{Pauli:1985ps,Brodsky:1997de,Heinzl:2000ht,Vary:2009gt,Brodsky:2014yha,Leitao:2017esb,Li:2017mlw}.
  By integrating out also the transverse momentum one obtains the parton distribution amplitude (PDA).

  In practice we employ a scalar model, namely the massive version of the Wick-Cutkosky model~\cite{Wick:1954eu,Cutkosky:1954ru,Nakanishi:1969ph}, which
  encapsulates the relevant features that are also present in more general situations and
  useful for testing the method. In particular, this model allows for detailed comparisons with the Nakanishi method
  and its application to LFWFs~\cite{Karmanov:2005nv,Frederico:2013vga}, and
  the results from both methods will turn out to be in excellent agreement.
  Moreover, the contour-deformation technique is general and not restricted to LFWFs, so it can be applied to hadron-to-hadron transition amplitudes as in Eq.~\eqref{2pt-correlator}
  and more general theories such as QCD in the future.

  The article is organized as follows. In Sec.~\ref{sec:minkowski} we establish the main formalism using Minkowski conventions
  and work out the LFWF for a simple monopole amplitude.
  In Sec.~\ref{sec:LF-euclidean} we derive the general expression for the LFWF in a Euclidean metric and analyze the singularity structure of the integrand.
  In Sec.~\ref{sec:bse} we calculate the LFWF dynamically from its Bethe-Salpeter equation using contour deformations, and we discuss the corresponding results.
  Sec.~\ref{sec:generalizations} deals with generalizations to unequal masses and complex propagator poles. We conclude in Sec.~\ref{sec:summary}.
  Two appendices provide details on the Nakanishi representation and the general properties of the singularities that appear in the integrands.

\newpage

\section{Light front in Minkowski space}\label{sec:minkowski}

\subsection{Definitions}\label{sec:defs}

      We begin with some basic definitions.
      In Minkowski conventions, one may define the light-front components of a four-vector $p^\mu$ by $p^\pm = p^0 \pm p^3$ such that
      \begin{equation}\label{lf-kinematics}
         p^3 = \frac{p^+-p^-}{2}\,, \quad
         p^0 = \frac{p^+ + p^-}{2} \,.
      \end{equation}
      A scalar product of two four-vectors then becomes
      \begin{equation}\label{sc-pr}
         k\cdot p = k^0 \,p^0 - \vect{k}\cdot\vect{p} = \frac{1}{2}\,(k^- p^+ + k^+ p^-) - \vect{k}_\perp \cdot \vect{p}_\perp \,,
      \end{equation}
      which implies $p^2 = p^+ p^- - \vect{p}_\perp^2$.
       For an onshell  particle with $p^2 = m^2$ and $p^0 > 0$, this entails $p^+ + p^- > 0$, $p^+ p^- = \vect{p}_\perp^2 + m^2 > 0$
      and therefore $p^\pm > 0$.
      The four-momentum integral in light-front variables reads
      \begin{equation}\label{lf-var-int}
         \int d^4p = \frac{1}{2} \int d^2\vect{p}_\perp \int dp^+ \int dp^-\,.
      \end{equation}

      We now consider the BSWF $\Psi(z,P)$ defined in Eq.~\eqref{bswf} for a bound state of two valence particles.
      We restrict ourselves to a scalar bound state of two scalar particles, although
      the generalization to more general theories such as QCD is straightforward.
      $P$ is the onshell total momentum ($P^2=M^2$) of the bound state and $z$ is the relative
      coordinate between the two constituents. The BSWF in momentum space is obtained by taking the Fourier transform,
        \begin{equation}\label{bswf-coordinate-space}
           \Psi(z,P) = \int\!\! \frac{d^4q}{(2\pi)^4}\,e^{-iq\cdot z}\,\Psi(q,P)\,,
        \end{equation}
         and differs from the BS amplitude $\Gamma(q,P)$  by attaching external propagator legs:
         \begin{equation}\label{bswf-mom-space}
            \Psi(q,P) = G_0(q,P)\,i\Gamma(q,P)\,.
         \end{equation}
         Here, $G_0(q,P)$ is the product of the particle propagators, e.g.,
         for tree-level propagators
         \begin{equation}\label{propagator-product}
            G_0(q,P) = \frac{i}{q_1^2-m_1^2+i\epsilon}\,\frac{i}{q_2^2-m_2^2+i\epsilon}\,,
         \end{equation}
         where $q$ is the relative momentum between the constituents (cf.~Fig.~\ref{fig:correlators}) and
        the two particle momenta are given by
         \begin{equation}\label{momenta-11}
            q_1 = q + \frac{1+\varepsilon}{2}P \,,  \qquad
            q_2 = -q + \frac{1-\varepsilon}{2}P\,.
         \end{equation}
         The variable $\varepsilon \in [-1,1]$ is an arbitrary momentum partitioning parameter and equal momentum partitioning corresponds to $\varepsilon = 0$.

         \pagebreak

        The light-front wave function (LFWF) $\psi(q^+,\vect{q}_\perp) $ is  defined as the Fourier transform of the BSWF restricted to $z^+ = 0$,
        which amounts to an integration over $q^-$ in momentum space:
        \begin{equation}\label{def-lfwf}
        \begin{split}
           \psi(q^+,\vect{q}_\perp) &= 2\mN P^+ \int d^4z\,e^{iq\cdot z}\,\Psi(z,P)\,\delta(z^+) \\
                        &= \mN P^+ \int_{-\infty}^\infty \frac{dq^-}{2\pi} \,\Psi(q,P)\,.
        \end{split}
        \end{equation}
        We used Eqs.~\eqref{sc-pr}, \eqref{lf-var-int} and \eqref{bswf-coordinate-space}, and $\mN$ is a normalization factor\footnote{We introduced $\mN$
        to account for different possible normalizations of the LFWF employed in the literature. Independently of this, the BS amplitude satisfies
        a canonical normalization condition, which is not important in what follows but implies that $\Gamma$ in the scalar theory has mass dimension 1.
        Here we consider $\Gamma$ to be dimensionless, such that the mass dimension enters through the factor $\mN$ and $\Psi$ has dimension $-4$.}
         with mass dimension 1.

        We now introduce a variable $\alpha \in [-1,1]$ through
        \begin{equation}\label{q-k-splitting}
           q = k + \frac{\alpha-\varepsilon}{2}\,P  \qquad \text{with}\quad k^+ = 0\,,
        \end{equation}
        such that the four-momenta in Eq.~\eqref{momenta-11} become
         \begin{equation}\label{momenta-12}
            q_1 = k + \frac{1+\alpha}{2}P \,,  \qquad
            q_2 = -k + \frac{1-\alpha}{2}P\,.
         \end{equation}
         The usual notation in terms of the longitudinal momentum fraction $\xi \in [0,1]$,
         \begin{equation}
             q_1^+ = \xi\,P^+\,, \qquad
             q_2^+ = (1-\xi)\,P^+\,,
         \end{equation}
         then follows from the identification
         \begin{equation}\label{xi}
            \xi = \frac{1+\alpha}{2}\,.
         \end{equation}
         In the following we set $\varepsilon=0$ for equally massive constituents ($m_1=m_2=m$), and
         we work with the variable $\alpha$ instead of $\xi$ to  allow for compact formulas where
         the symmetry $\alpha \to -\alpha$ is manifest.
         Correspondingly, we write the LFWF as
        \begin{equation}\label{def-lfwf-2}
           \psi(\alpha,\vect{k}_\perp) =  \mN P^+ \int_{-\infty}^\infty \frac{dq^-}{2\pi} \,\Psi(q,P)\big|_{q^+ = \frac{\alpha}{2} P^+, \,\vect{\scriptstyle q}_\perp = \vect{\scriptstyle k}_\perp} \,.
        \end{equation}

        The PDA $\phi(\alpha)$ and distribution function $u(\alpha)$ then follow from
        an integration over $d^2\vect{k}_\perp$:
        \begin{equation}\label{da-df}
        \begin{split}
            \phi(\alpha) &= \frac{1}{16\pi^3 f} \int d^2\vect{k}_\perp\,\psi(\alpha,\vect{k}_\perp)\,, \\
            u(\alpha) &= \int d^2\vect{k}_\perp\,|\psi(\alpha,\vect{k}_\perp)|^2\,.
        \end{split}
        \end{equation}
        The decay constant $f$ is proportional to the integrated BSWF;
        integrating $\phi(\alpha)$ over $\alpha$, one finds
        \begin{equation}
           \frac{f}{2} \int_{-1}^1 d\alpha\,\phi(\alpha)  =  \mN \int \!\! \frac{d^4q}{(2\pi)^4}\,\Psi(q,P)\,.
        \end{equation}

\subsection{Light-front wave function for monopole}\label{sec:lfwf-monopole}

        It is illustrative to work out the LFWF
        for the  case where the BS amplitude is a simple monopole,
        \begin{equation}\label{monopole}
           \Gamma(q,P) =  -\frac{m^2}{q^2 -m^2\gamma +i\epsilon} \quad \text{with} \quad \gamma > 0\,,
        \end{equation}
        which is easily evaluated using residue calculus.
        In the rest frame of the bound state one has
        \begin{equation}
           P = \left[ \begin{array}{c} M \\ \vect{0} \\ 0 \end{array}\right], \qquad
           q = \left[ \begin{array}{c} q^0 \\ \vect{q}_\perp \\ q^3 \end{array}\right],
        \end{equation}
        where $P^+ = P^- = M$ and thus $q^+ = \alpha M/2$.
        The  bound-state mass $M$ must be below the two-particle threshold ($M<2m$), and
        to eliminate the mass parameter $m$ from the equations we define
        \begin{equation}\label{LIs}
           t = -\frac{M^2}{4m^2}, \quad q^- = \frac{2m^2}{P^+} w\,, \quad \vect{q}_\perp^2 = m^2 x
        \end{equation}
        with $t \in [-1,0]$, $w \in \mathds{R}$ and $x>0$.
        This entails
        \begin{equation}
        \begin{split}
           q^2 &= m^2\,(\alpha w - x), \\
           q\cdot P &= m^2\,(w-\alpha t),\\
           q_1^2  &= m^2\left[ (\alpha+1)(w-t) - x\right], \\
           q_2^2  &= m^2\left[ (\alpha-1)(w+t) - x\right].
        \end{split}
        \end{equation}

         As a result, the LFWF in Eq.~\eqref{def-lfwf} and  BSWF in~(\ref{bswf-mom-space}--\ref{propagator-product}) take the form
        \begin{align}
           \psi(\alpha,x) &= \frac{\mN m^2}{\pi} \int\limits_{-\infty}^\infty dw \,\Psi(q,P)\,,  \label{lfwf-3}  \\
           \Psi(q,P) &= \frac{1}{im^4}\,\frac{1}{\alpha\,(1-\alpha^2)}  \,\frac{1}{w-w_+}\,\frac{1}{w-w_-}\,\frac{1}{w-w_0}\,, \nonumber
        \end{align}
         where $w=w_\pm$ are the propagator pole locations corresponding to $q_{1,2}^2 = m^2-i\epsilon$ and $w_0$ is the pole from the BS amplitude:
         \begin{equation}\label{pole-positions-w}
            w_\pm = \pm \left( t + \frac{x+1-i\epsilon}{1\pm \alpha}\right), \quad
            w_0 = \frac{x+\gamma-i\epsilon}{\alpha} \,.
         \end{equation}
         The residues of the BSWF at the three poles, multiplied with $2\pi i$, are
         \begin{equation}
         \begin{split}
            \mR_+ &= \frac{2\pi}{m^4}\,\frac{1}{\alpha\,(1-\alpha^2)} \frac{1}{w_+-w_-}\,\frac{1}{w_+-w_0}\,, \\
            \mR_- &= \frac{2\pi}{m^4}\,\frac{1}{\alpha\,(1-\alpha^2)} \frac{1}{w_--w_+}\,\frac{1}{w_--w_0}\,, \\
            \mR_0 &= \frac{2\pi}{m^4}\,\frac{1}{\alpha\,(1-\alpha^2)} \frac{1}{w_0-w_+}\,\frac{1}{w_0-w_-}\,.
         \end{split}
         \end{equation}
         Using Eq.~\eqref{pole-positions-w} this yields
         \begin{equation}
         \begin{split}
            \mR_\pm &= \mp \frac{\pi}{m^4}\,\frac{1}{x+A}\,\frac{1\pm \alpha}{x+A+ (1 \pm \alpha)\, B}\,, \\
            \mR_0 &=  \frac{\pi}{m^4}\,\frac{2\alpha}{(x+A+B)^2 - \alpha^2 B^2}
         \end{split}
         \end{equation}
         with
         \begin{equation}
            A =1+(1-\alpha^2)\,t\,,  \qquad
            B = \gamma-1-t \,.
         \end{equation}
         One may verify that $\mR_++\mR_-+\mR_0=0$, i.e., the sum of the residues vanishes as it should.

  \begin{figure}[t]
  \includegraphics[width=0.85\columnwidth]{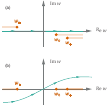}
  \caption{Sketch of the singularities in the complex $w$ plane for $0 < \alpha < 1$. (a) is the situation in Eqs.~(\ref{lfwf-3}--\ref{pole-positions-w}) with
            $i\epsilon$ in the denominators, whereas (b) corresponds to Eq.~\eqref{lfwf-3a} with $i\epsilon$ in the integration path.
            For $|\alpha|<1$ the results are identical, but for general values of $\alpha$ only (b) leads to an analytic function.}
  \label{fig-int-path}
  \end{figure}

\begin{figure}[t]
  \includegraphics[width=0.8\columnwidth]{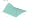}
  \caption{Light-front wave function~\eqref{lfwf-result-monopole} for the monopole model with $t=-0.5$, $\gamma=2$ and normalized to $\psi(0,0)$.}
  \label{fig-monopole}
  \end{figure}

        From Eq.~\eqref{pole-positions-w},
         the propagator poles at $w=w_+(w_-)$ in the complex $w$ plane lie  below (above) the real axis.
         The displacement of the $w_0$ pole depends on  sign$(\alpha)$; for $\alpha>0$ it lies below the real axis and for $\alpha<0$ above.
         This is sketched in Fig.~\ref{fig-int-path}(a) for $\alpha >0$, where the horizontal tracks show the movement of the poles for $w_\pm \pm \lambda$ and $w_0 + \lambda$ with a parameter $\lambda > 0$.
         Thus, when integrating $w$ along the real axis and closing the contour at complex infinity,
         one picks up the residue $\mR_-$ for $\alpha>0$ and $\mR_-+\mR_0 = -\mR_+$ for $\alpha < 0$.
         The combined result for the LFWF is therefore
        \begin{equation}\label{lfwf-result-monopole}
           \psi(\alpha,x) = \frac{\mN}{m^2}\,\frac{1}{x+A}\,\frac{1-|\alpha|}{x+A+(1-|\alpha|)\, B}\,,
        \end{equation}
        which is shown in Fig.~\ref{fig-monopole} for exemplary values of $t$ and $\gamma$.

         Abbreviating $C=(1-|\alpha|)\,B/A$, the corresponding distribution amplitude $\phi(\alpha)$ in Eq.~\eqref{da-df} is
        \begin{equation}
           \phi(\alpha) = \frac{m^2}{(4\pi)^2 f} \int\limits_0^\infty dx\,\psi(\alpha,x)
                        = \frac{\mN}{(4\pi)^2 f}\,\frac{\ln(1+C)}{B}
        \end{equation}
        and the distribution function $u(\alpha)$ reads
        \begin{align}
           u(\alpha) &= \pi m^2 \int\limits_0^\infty dx\,|\psi(\alpha,x)|^2 \\
                        &= \frac{\pi\, |\mN|^2}{m^2}\,\frac{1}{AB^2}\left[1+\frac{1}{1+C}-\frac{2}{C}\,\ln(1+C)\right]. \nonumber
        \end{align}
        For $\gamma=1+t$ and thus $B=0$, and suppressing the prefactors, these quantities reduce to
        \begin{equation}
        \psi(\alpha,x)  \propto \frac{1-|\alpha|}{(x+A)^2}\,, \qquad
        \begin{array}{rl}
           \phi(\alpha) & \displaystyle \propto \frac{1-|\alpha|}{A}\,, \\[3mm]
           u(\alpha) & \displaystyle \propto \frac{(1-|\alpha|)^2}{3A^3}\,.
        \end{array}
        \end{equation}

        Strictly speaking, with the $i\epsilon$ factors as in Eq.~\eqref{pole-positions-w} these results are only valid
        for real values of $\alpha$ with $|\alpha|<1$. For $|\alpha|>1$, all poles in the complex $w$ plane in Fig.~\ref{fig-int-path}(a) move
        either to the upper or lower half plane, shifted by $i\epsilon$, such that the closed contour is empty and the resulting LFWF is zero.
        Thus, the LFWF only has support for $-1 < \alpha < 1$.

\begin{figure*}[t]
  \includegraphics[width=0.8\textwidth]{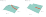}
  \caption{Light-front wave function~\eqref{lfwf-result-monopole} for the monopole in the complex $\alpha^2$ plane.
           For the integration path in Fig.~\ref{fig-int-path}(a) the function would only have support for $\alpha^2>0$ (thick orange curves)
           but vanish everywhere else. }
  \label{fig-monopole-complex}
  \end{figure*}

        On the other hand,
         the $i\epsilon$ prescription has its origin in the imaginary-time boundary conditions, which
          translates to analogous boundary conditions for the $q^-$ and $w$ integration:
        \begin{equation}\label{lfwf-3a}
           \psi(\alpha,x) = \frac{\mN m^2}{\pi} \int\limits_{-\infty(1+i\epsilon)}^{\infty(1+i\epsilon)} dw \,\Psi(q,P)\,.
        \end{equation}
        This suggests an integration path like  in Fig.~\ref{fig-int-path}(b), which starts at a finite imaginary part $\text{Im}\,w = -\infty \cdot \varepsilon$
        below all poles in the integrand and ends at $\text{Im}\,w = +\infty \cdot \varepsilon$ above all poles.
        For $-1 < \alpha < 1$ the results from (a) and (b) are identical,
        but for $|\alpha|>1$ or for complex values of $w_\pm$ and $w_0$ they are different.
        In particular, only (b) leads to an analytic function but (a) does not.
        Substituting $|\alpha| \to \sqrt{\alpha^2}$ in Eq.~\eqref{lfwf-result-monopole}
         yields the result of (b) for any $\alpha$, $x$, $t$, $\gamma \in \mathds{C}$,
        which is an analytic function in the complex $\alpha^2$ plane and plotted in Fig.~\ref{fig-monopole-complex}.

  \begin{figure*}[t]
  \includegraphics[width=0.8\textwidth]{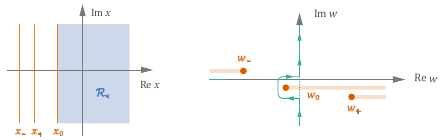}
  \caption{Different regions in the complex $x$ plane separated by the branch cuts~\eqref{xcut-M}.
           To obtain the correct light-front wave function outside $\mR_\epsilon$, one would need to deform the Euclidean integration contour.}
  \label{fig-poles-in-x}
  \end{figure*}

        By contrast, for complex values of $\alpha$ (with $t$, $x$, $\gamma$ real) Eq.~\eqref{pole-positions-w} implies
        \begin{align}
           \text{Im}\,w_\pm &= \pm \text{Im}\,\frac{x+1-i\epsilon}{1\pm \alpha} = \frac{(x+1)\,\text{Im}\,\alpha^\ast \mp \epsilon\,(1\pm \text{Re}\,\alpha)}{|1\pm\alpha|^2}\,, \nonumber \\
           \text{Im}\,w_0 &= \text{Im}\,\frac{x+\gamma-i\epsilon}{\alpha} = \frac{(x+\gamma)\,\text{Im}\,\alpha^\ast - \epsilon\,\text{Re}\,\alpha}{|\alpha|^2}\,.
        \end{align}
        For $\text{Im}\,\alpha \neq 0$, the $i\epsilon$ factors become irrelevant and for $x+1>0$ and $x+\gamma>0$
        the three poles always lie in the same half plane.
        Also for $\text{Im}\,\alpha = 0$ and $|\alpha|>1$ the three poles fall in the same half plane.
        Therefore,
        the LFWF from (a) vanishes everywhere except for $\alpha^2 \in \mathds{R}_+$ and is thus not an analytic function.

        In the following we adopt the interpretation (b) for the $i\epsilon$ prescription, since this
        is what generates an analytic function and will allow us to perform analytic continuations.
        As long as the integrand vanishes sufficiently fast at complex infinity,
        one can then equivalently perform a Wick rotation and integrate $w$ along a Euclidean integration path from $-i\infty$ to $+i\infty$.

         To facilitate the numerical treatment,  instead of working out the poles in the complex $w$ plane we consider the respective pole positions
         in the complex $x$ plane by solving Eq.~\eqref{pole-positions-w} for $x$:
         \begin{equation}\label{xcut-M}
         \begin{split}
             x_\pm &= (1\pm\alpha)(\pm w - t) - 1\,, \\
             x_0 &= \alpha w - \gamma\,.
         \end{split}
         \end{equation}
         When integrating $w \in (-i\infty,i\infty)$, the pole positions turn
         into branch cuts in $x \in \mathds{C}$ as sketched in the left panel of Fig.~\ref{fig-poles-in-x}. They are defined by
         \begin{equation}\label{xcut-M2}
         \begin{split}
             \text{Re}\,x_\pm^\text{cut} &= - t\,(1\pm\alpha) - 1\,, \\
             \text{Re}\,x_0^\text{cut} &=- \gamma\,.
         \end{split}
         \end{equation}
         These cuts separate different regions in the complex $x$ plane with different values of the integral.
         The region corresponding to the proper $i\epsilon$ prescription is shown in blue and
         in the following we refer to it as $\mR_\epsilon$. A calculation of the LFWF inside this region
         returns the correct result in Eq.~\eqref{lfwf-result-monopole}.

         Vice versa, for the alignment displayed in Fig.~\ref{fig-poles-in-x}, $x$ behind the first cut corresponds to the situation where the $w_0$ pole
         has moved to the wrong side of the Euclidean integration contour in the complex $w$ plane,
         thereby not summing the correct residues and giving the wrong value of the integral. If we wanted to know the LFWF for $x$ behind the first cut,
         we would need to deform the Euclidean contour in $w$ (right panel in Fig.~\ref{fig-poles-in-x}).
         We refrain from doing so in what follows but instead restrict the calculations to the region $\mR_\epsilon$.

         Below we will transform Eq.~\eqref{lfwf-3} to hyperspherical variables, in which case
         the resulting branch cuts form more complicated curves
         but  still define a corresponding $\mR_\epsilon$ region.
         The LFWF can then be calculated numerically inside that region.
         In turn, this region may not include the whole positive real axis in $x$, which requires contour deformations in the complex $x$ plane when
         calculating the light-front distributions in Eq.~\eqref{da-df} or when solving for the BSWF $\Psi(q,P)$.

      \section{Light front in Euclidean space}\label{sec:LF-euclidean}

\subsection{Euclidean conventions}

        The goal in the following is to transfer Eq.~\eqref{def-lfwf} to a Euclidean metric $(+,+,+,+)$.
        A Euclidean four-vector is defined by
        \begin{equation}
           a_E = \left[ \begin{array}{c} \vect{a} \\ a_4 \end{array}\right] = \left[ \begin{array}{c} \vect{a} \\ ia^0 \end{array}\right].
        \end{equation}
        In the Euclidean metric the distinction between upper and lower components becomes irrelevant, and scalar products of four-vectors pick up minus signs: $a_E \cdot b_E = -a\cdot b$.
        The light-front variables in Eq.~\eqref{lf-kinematics} are independent of the metric, so one has $p^\pm = -ip_4 \pm p_3$ and
      \begin{equation}
         p_3 = \frac{p^+-p^-}{2}\,, \quad
         p_4 = i\,\frac{p^+ + p^-}{2} \,.
      \end{equation}
      The relation~\eqref{sc-pr} turns into
            \begin{equation}\label{sc-pr-2}
         k_E\cdot p_E = \vect{k}_\perp \cdot \vect{p}_\perp  - \frac{1}{2}\,(k^- p^+ + k^+ p^-) \,.
      \end{equation}

      We also take the opportunity to remove the various signs and $i$ factors appearing in the Minkowski quantities.
      In the Euclidean notation, a scalar tree-level propagator is $1/(q^2+m^2)$,
      which entails $G_0^E = -G_0^M$ for the propagator product~\eqref{propagator-product}.
      With $\Gamma_E = \Gamma_M$ for the BS amplitude and $\Psi_E = i\Psi_M$ for the BSWF,
      Eq.~\eqref{bswf-mom-space} becomes $\Psi_E = G_0^E\,\Gamma_E$.
      We now drop the label `E' and continue to work with Euclidean conventions.

\subsection{Light-front wave function}

      To work out the LFWF in Euclidean kinematics, we start from Eq.~\eqref{def-lfwf-2}:
        \begin{equation}\label{def-lfwf-3}
           \psi(\alpha,\vect{k}_\perp) =  \mN P^+ \int_{-i\infty}^{i\infty} \frac{dq^-}{2i\pi} \,\Psi(q,P)\big|_{q^+ = \frac{\alpha}{2} P^+, \,\vect{\scriptstyle q}_\perp = \vect{\scriptstyle k}_\perp} \,.
        \end{equation}
      Because the light-front variables are independent of the metric, the formula has the same form as earlier
      except that the integration over $q^-$ now proceeds from $-i\infty$ to $+i\infty$, and the factor $i$
      in the denominator comes from the Euclidean definition of the BSWF $\Psi(q,P)$ as mentioned above. The latter is given by
         \begin{equation}\label{bswf-mom-space-E}
         \begin{split}
            \Psi(q,P) &= G_0(q,P)\,\Gamma(q,P)\,, \\
            G_0(q,P) &= \frac{1}{q_1^2+m^2}\,\frac{1}{q_2^2+m^2}\,,
         \end{split}
         \end{equation}
         where the corresponding BS amplitude $\Gamma(q,P)$ is the dynamical solution of the BSE which we will discuss in Sec.~\ref{sec:bse}.
         For the monopole example~\eqref{monopole} it is given by
        \begin{equation}\label{monopole-E}
           \Gamma(q,P) =  \left( \frac{q^2}{m^2} + \gamma\right)^{-1}.
        \end{equation}
        We note that for a scalar bound state with scalar constituents, all quantities $\Psi(q,P)$, $\Gamma(q,P)$ and $G_0(q,P)$ are Lorentz-invariant.

       Like in Eq.~\eqref{q-k-splitting}, we define a four-momentum $k$ through $q=k+\alpha P/2$,
       where for now we set $\varepsilon=0$ for equally massive constituents.
       Because the BSWF is Lorentz-invariant, it can only depend on the Lorentz invariants $k^2$, $k\cdot P$ and $P^2 = -M^2$ together with the momentum partitioning $\alpha$.
       We express them through the dimensionless variables $x$, $\omega$ and $t$:
         \begin{equation}\label{kin-x-omega-t}
            k^2 = m^2 x\,, \quad
            k\cdot P = 2m^2\sqrt{xt}\,\omega\,, \quad
            P^2 = 4m^2\,t\,.
         \end{equation}
      The self-consistent domain of the BSE solution in Sec.~\ref{sec:bse}
      is $x>0$ and $\omega \in [-1,1]$, where $\omega = \hat{k}\cdot\hat{P}$ is the cosine of a four-dimensional angle (a hat denotes a unit four-vector).
       The Lorentz invariants $q^2$ and $q\cdot P$ then become
       \begin{equation}\label{q2-qP-12}
       \begin{split}
          \frac{q^2}{m^2} &= x + \alpha^2\,t + 2\alpha\sqrt{xt}\,\omega \,, \\
          \frac{q\cdot P}{m^2} &= \frac{k\cdot P}{m^2} + 2\alpha t = 2\sqrt{xt}\,\omega + 2\alpha t \,.
       \end{split}
      \end{equation}

         In Euclidean conventions the four-momenta are conveniently expressed in hyperspherical variables, which are more closely related to the Lorentz invariants of the system.
         In a general moving frame of the total momentum $P$, this amounts to
         \begin{equation}\label{alternative-kinematics}
         \begin{split}
           k &= m\sqrt{x}\left[ \begin{array}{l}\sqrt{1-z^2}\sqrt{1-y^2} \, \sin\psi \\ \sqrt{1-z^2}\sqrt{1-y^2} \,\cos\psi \\ \sqrt{1-z^2}\,y \\ z \end{array}\right],
           \\
           P &= 2m\sqrt{t} \left[ \begin{array}{c} 0 \\ 0 \\ \sqrt{1-Z^2} \\ Z \end{array}\right].
         \end{split}
        \end{equation}
        When the BSE is solved in the moving frame, % to obtain the BSWF $\Psi(q,P)$,
        the natural domain of these variables is $x>0$, $\psi\in[0,2\pi)$ and  $z, y, Z \in [-1,1]$.
        Because we can always choose $\vect{P}_\perp=0$, $\vect{q}_\perp = \vect{k}_\perp$ is automatic.
        On the other hand, the condition $k^+ = 0$ in light-front kinematics entails
        \begin{equation}\label{lf-y}
            y = \frac{iz}{\sqrt{1-z^2}} \quad \Rightarrow \quad           \sqrt{1-z^2}\sqrt{1-y^2}=1 \,,
        \end{equation}
        so that the Lorentz-invariant variable $x = k^2/m^2 = \vect{k}_\perp^2/m^2 = \vect{q}_\perp^2/m^2$
        assumes the meaning of  the squared transverse momentum.

            \begin{figure*}[t]
                    \begin{center}
                    \includegraphics[width=0.97\textwidth]{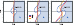}
                    \caption{Exemplary cut configurations in the complex $\sqrt{x}$ plane for $\alpha=0.6$, $\gamma=4$ and three different values of $\sqrt{t}$.
                             The three lines correspond to the propagator cuts $\sqrt{x}_\pm$ (blue, red) and the cut $\sqrt{x}_0$ from the monopole amplitude  (orange)  in Eq.~\eqref{cuts-lfwf-0}.
                             The region $\mR_\epsilon$ shown in blue connects the origin $\sqrt{x}=0$ with $\sqrt{x}\to\infty$.}\label{fig:cuts-7}
                    \end{center}
                    \vspace{-5mm}
            \end{figure*}

       Now let us transform the integration over $dq^-$  in Eq.~\eqref{def-lfwf-3} to Euclidean variables.
       From Eq.~\eqref{sc-pr-2} together with $\vect{P}_\perp=0$, $q^+ = \alpha P^+/2$ and $P^+P^- = M^2$, we have
       \begin{equation}
          \frac{q\cdot P}{m^2} = -\frac{1}{2m^2}\left( q^- P^+ + q^+ P^-\right) = -\frac{q^- P^+}{2m^2} + \alpha t\,,
       \end{equation}
       and comparison with Eqs.~\eqref{q2-qP-12} and~\eqref{LIs} gives
       \begin{equation}\label{q-euclid}
          q^- = -\frac{2m^2}{P^+}\left( 2\sqrt{xt}\,\omega + \alpha t\right) \; \Rightarrow \;
          \omega = -\frac{w+\alpha t}{2\sqrt{xt}}\,.
       \end{equation}
       For given values of $x>0$, $\alpha \in [-1,1]$  and $t\in[-1,0]$, the square root $\sqrt{t} = iM/(2m)$ implies that the integration contour $q^- = -i\infty \dots +i\infty$ maps to
       $\omega = +\infty \dots -\infty$, so we arrive at
        \begin{equation}\label{def-lfwf-12}
           \psi(\alpha,x,t) = \frac{\mN m^2}{i\pi} \,2\sqrt{xt}  \, \int\limits_{-\infty}^\infty d\omega \,\Psi(x,\omega,t,\alpha)\big|_{k^+ = 0} \,.
        \end{equation}
        Here we also made the dependence of the BSWF on the Lorentz invariants $x$, $\omega$, $t$ and $\alpha$ explicit.
        The endpoints of the Euclidean integration contour are $\omega = \pm\infty$, but the actual contour
        depends on the singularities of the integrand which we will determine below.
%        \footnote{We
%        dropped the constant term $\alpha t$ from Eq.~\eqref{q-euclid} in the integration path,
%        as this would merely lead to different branch cuts and thus a different $\mR_\epsilon$ region as discussed below.}

        What does the constraint $k^+ = 0$ mean for the BSWF?
        Eqs.~(\ref{alternative-kinematics}--\ref{lf-y}) entail
         \begin{equation}
            \omega = zZ + y\sqrt{1-z^2}\sqrt{1-Z^2}
                   = z\,( Z + i\sqrt{1-Z^2})\,,
        \end{equation}
         and from $P^+ = -iP^4 + P^3$ one finds $\omega = z P^+/M$.
%         \begin{equation}
%           \omega  = z\,\frac{P^+}{M}\,.
%         \end{equation}
         This does not pose any additional constraints on $\omega$; e.g., for $P^+ > 0$ and $z\in[-1,1]$ also the domain $\omega \in \mathds{R}$ remains the same as before.
         Then, because the BSWF is frame-independent, we can calculate it in any frame and
         instead of Eq.~\eqref{alternative-kinematics} we may as well work in the rest frame of the bound state:
         \begin{equation}\label{rest-frame}
           k = m\sqrt{x}\left[ \begin{array}{c} 0 \\ 0 \\ \sqrt{1-\omega^2}  \\ \omega \end{array}\right], \quad
           P = 2m\sqrt{t} \left[ \begin{array}{c} 0 \\ 0 \\ 0 \\ 1 \end{array}\right].
        \end{equation}
        Formally, the conditions $k^+ = 0$ and $\vect{k}_\perp^2 = m^2 x$ are not meaningful in this frame,
        but the Lorentz invariance of the BSWF implies that the result must be identical to that obtained in a moving frame if those conditions are imposed.
        As a consequence, the final expression for the LFWF becomes
        \begin{equation}\label{def-lfwf-13}
           \psi(\alpha,x,t) = \frac{\mN m^2}{i\pi} \,2\sqrt{xt}  \, \int\limits_{-\infty}^\infty d\omega \,\Psi(x,\omega,t,\alpha)\,.
        \end{equation}
        Note that this formula  no longer makes any reference to light-front variables but only depends on Lorentz-invariant quantities.
        From Eqs.~\eqref{momenta-11} and~\eqref{momenta-12} one can see that in practice it amounts to
        evaluating the BSWF for a general momentum partitioning parameter $\alpha$ and subsequently integrating over $\omega$.

        The distribution amplitude $\phi(\alpha)$ and distribution function $u(\alpha)$ in Eq.~\eqref{da-df} then follow accordingly:
        \begin{equation}\label{da-df-2}
        \begin{split}
            \phi(\alpha) &= \frac{m^2}{(4\pi)^2 f} \int_0^\infty dx\,\psi(\alpha,x,t)\,, \\
            u(\alpha) &= \pi m^2 \int_0^\infty dx\,|\psi(\alpha,x)|^2\,.
        \end{split}
        \end{equation}

        For  general theories and bound states with arbitrary spin, $\Psi$ is not Lorentz invariant but still
          \textit{covariant}; thus one can expand it in a Lorentz-covariant tensor basis with Lorentz-invariant coefficients.
        The operations in light-front kinematics ($k^+=0$, taking the $\gamma^+$ component, etc.) then affect the tensor basis
        but still  leave the dressing functions invariant. In this way Eq.~\eqref{def-lfwf-13} can also be generalized to
        matrix elements of the form $\langle P_f | \mathsf{T}\,\Phi(z)\,\mathcal{O}\,\Phi(0) | P_i\rangle$
         taken on the light front,
        which enter in the definition of parton distributions such as PDFs, GPDs and TMDs.

        In going from Eq.~\eqref{lfwf-3a} to~\eqref{def-lfwf-13}  we effectively changed the integration variable
        from $w$ to $\omega$ through Eq.~\eqref{q-euclid}. This
        also changes the branch cuts in the complex $x$ plane and the corresponding $\mR_\epsilon$ regions, which
        are illustrated in Fig.~\ref{fig:cuts-7} and discussed in detail in the following.
        This is also the reason that allowed us to drop the constant term $\propto \alpha t$ from Eq.~\eqref{q-euclid} in the integration path over $\omega$,
        as this would only lead to different branch cuts and thus a different $\mR_\epsilon$ region.

\subsection{Singularity structure}

        The difficulty in evaluating Eq.~\eqref{def-lfwf-13} is the singularity structure of $\Psi$.
             On  one hand, the propagators in Eq.~\eqref{bswf-mom-space-E} (and generalizations thereof) induce singularities in $G_0$; on the other hand, the  solution of the
             BSE may produce singularities in the BS amplitude $\Gamma$.

         To determine the singularity structure in the complex $\sqrt{x}$ plane, we consider again the tree-level propagators in Eq.~\eqref{bswf-mom-space-E}
         and the monopole amplitude~\eqref{monopole-E}.
         The momenta of the propagators are given in Eq.~\eqref{momenta-12}, which together with~\eqref{kin-x-omega-t} entails
         \begin{equation}\label{kinematics-E2}
         \begin{split}
            \frac{q_{1,2}^2}{m^2} +1 &= x - 2\sqrt{x}\, \omega A_\pm + A_\pm^2  + 1\,, \\
            \frac{q^2}{m^2} + \gamma &= x - 2\sqrt{x}\, \omega A_0 + A_0^2  + \gamma\,,
         \end{split}
         \end{equation}
         with $A_\pm = \mp (1\pm\alpha) \sqrt{t}$ and $A_0 = - \alpha \sqrt{t}$.
         These expressions have zeros at
       \begin{equation}
       \begin{split}
          \sqrt{x}_\pm &= A_\pm \left[ \omega + i\lambda\sqrt{1-\omega^2 + \frac{1}{A_\pm^2}}\,\right],  \\
          \sqrt{x}_0 &= A_0 \left[ \omega + i\lambda\sqrt{1-\omega^2 + \frac{\gamma}{A_0^2}}\right], \label{cuts-lfwf-0}
       \end{split}
       \end{equation}
       where $\lambda = \pm 1$ denotes the two solution branches in each case.
       For real values of $\omega$, Eq.~\eqref{cuts-lfwf-0} describes branch cuts in the complex $\sqrt{x}$ plane which are shown in Fig.~\ref{fig:cuts-7}.
       They separate eight regions, which relate to the four distinct possibilities of picking up the three pole residues
       when integrating over $\omega$: picking up none of the residues is equivalent to picking up all three of them and gives zero;
       picking up one of them is equivalent to picking up the other two with the opposite direction of the integration contour.
       Without loss of generality, we can restrict ourselves to values of $\sqrt{t}=iM/(2m)$ in the upper right quadrant.
       The region $\mR_\epsilon$ corresponding to the proper $i\epsilon$ prescription is then the one shown in blue and connects the origin $\sqrt{x}=0$ with $\sqrt{x}\to\infty$.

       For the monopole example and $\sqrt{x} \in \mR_\epsilon$, one can easily verify that the numerical integration in Eq.~\eqref{def-lfwf-13} in combination with~(\ref{bswf-mom-space-E}--\ref{monopole-E})
       and~\eqref{kinematics-E2} reproduces the earlier  result for $\psi(\alpha,x,t)$ obtained  in Minkowski space, Eq.~\eqref{lfwf-result-monopole}.
       Note in particular that the vanishing of the LFWF at the endpoints $\alpha = \pm 1$ is automatic.

       The branch cuts in Fig.~\ref{fig:cuts-7} can also cross the real $\sqrt{x}$ axis, in which case it is
       no longer straightforward to compute the light-front distributions~\eqref{da-df-2} numerically
       and one must deform the integration contour in $\sqrt{x}$.
       The possible shapes of the functions in Eq.~\eqref{cuts-lfwf-0} are discussed in Appendix~\ref{app:cuts};
       from there it follows that for $\alpha\in[-1,1]$ and $\gamma > 0$, the cuts do not cross the real axis if
       \begin{equation}\label{constraint-1}
          |\sqrt{t}| < \text{min} \left[ \frac{1}{1\pm\alpha}\,, \, \frac{\sqrt{\gamma}}{|\alpha|} \right] \to \text{min} \left[ \frac{1}{2}\,, \, \sqrt{\gamma}\right],
       \end{equation}
       in which case no contour deformation is necessary.

       If $\sqrt{t}$ moves to the imaginary axis but does not satisfy this constraint,
       the branch cuts become increasingly circular.   Imaginary $\sqrt{t}$ with $0 < \text{Im}\sqrt{t} < 1$ corresponds to
       the physical situation $0<M<2m$ for bound states.
        In this case the only integration contour starting from the origin also coincides with the imaginary axis,
        which makes a numerical integration impossible.
        The strategy is then to compute the quantities of interest for complex values of $\sqrt{t}$ and take the limit $\text{Re}\sqrt{t}\to 0$ in the end.

            \begin{figure}[t]
                    \begin{center}
                    \includegraphics[width=0.85\columnwidth]{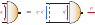}
                    \caption{Bethe-Salpeter equation.}\label{fig:bse}
                    \end{center}
                    \vspace{-5mm}
            \end{figure}

      \section{Dynamical calculation of light-front wave functions}   \label{sec:bse}

      \subsection{Bethe-Salpeter equation}

       In general, the BS amplitude $\Gamma(q,P) = \Gamma(x,\omega,t,\alpha)$
       is not available in a closed form but emerges dynamically from the solution of the homogeneous BSE, which is illustrated in Fig.~\ref{fig:bse} and reads
       \begin{equation}\label{bse-general}
          \Gamma(q,P) = \int \!\!  \frac{d^4q'}{(2\pi)^4} \,K(q,q',P)\,G_0(q',P)\,\Gamma(q',P)\,.
       \end{equation}
       $G_0(q,P)$ is the propagator product from Eq.~\eqref{bswf-mom-space-E} and $K(q,q',P)$ is
        the one-boson exchange kernel for a scalar particle with mass $\mu$,
       \begin{equation}
           K(q,q',P) = \frac{g^2}{(q-q')^2 + \mu^2} \,.
       \end{equation}
       Because  $g$ in the scalar theory is dimensionful, it is convenient to define a dimensionless coupling constant $c$
       and the mass ratio $\beta$:
       \begin{equation}\label{c-beta}
          c = \frac{g^2}{(4\pi m)^2}\,, \qquad \beta = \frac{\mu}{m}\,.
       \end{equation}
       Details on the solution of the scalar BSE can be found in Ref.~\cite{Eichmann:2019dts};
       the practical complication in our present case is the appearance of the momentum partitioning $\alpha \neq 0$.

       To this end, we set $q = k + \alpha P/2$ as before, and with $q' = k' + \alpha P/2$ also for the loop momentum we have $q-q' = k-k'$.
       We work in the rest frame defined by Eq.~\eqref{rest-frame} and
       \begin{equation}
           k' = m\sqrt{x'}\left[ \begin{array}{l} \sqrt{1-{\omega'}^2}\sqrt{1-{y'}^2}\,\sin\theta \\ \sqrt{1-{\omega'}^2}\sqrt{1-{y'}^2}\,\cos\theta \\ \sqrt{1-{\omega'}^2}\,y'  \\ \omega' \end{array}\right].
        \end{equation}
       Writing $\Gamma$, $K$ and $G_0$ in terms of Lorentz invariants,
       the kernel becomes
       \begin{equation}\label{kernel}
       \begin{split}
           K(x,x',\Omega) &= \frac{(4\pi)^2\,c}{x+x'+\beta -2\sqrt{xx'}\,\Omega}\,, \\[1mm]
           \Omega &= \hat{k}\cdot\hat{k}' = \omega\omega' + y \sqrt{1-\omega^2}\sqrt{1-{\omega'}^2}\,,
       \end{split}
       \end{equation}
       and the propagator product from Eqs.~\eqref{bswf-mom-space-E} and~\eqref{kinematics-E2} which carries the $\alpha$ dependence is
       \begin{align}
             &G_0(x,\omega,t,\alpha) = \frac{1}{m^4} \times \label{prop-matrix} \\
             &\times \frac{1}{\left[x + 1 + (1 + \alpha^2)\,t + 2\alpha\sqrt{xt}\,\omega\right]^2 -4t\left[\sqrt{x}\,\omega + \alpha\sqrt{t}\right]^2} .      \nonumber
       \end{align}
       Shifting the integration from $d^4q'$ to $d^4k'$, the integration measure is
       \begin{equation}
          \int d^4k' = \frac{m^4}{2} \int_0^\infty dx'\,x' \int_{-1}^1 d\omega'\sqrt{1-{\omega'}^2} \int_{-1}^1  dy' \int_{0}^{2\pi} d\theta
       \end{equation}
       and the BSE becomes
       \begin{equation}\label{bse-final}
       \begin{split}
          &\Gamma(x,\omega,t,\alpha) = \frac{m^4}{(2\pi)^3}\,\frac{1}{2}\int_0^\infty dx'\,x' \int_{-1}^1 d\omega'\sqrt{1-{\omega'}^2} \\[-1mm]
          & \quad \times G_0(x',\omega',t,\alpha) \int_{-1}^1  dy \,K(x,x',\Omega)\,\Gamma(x',\omega',t,\alpha)\,.
       \end{split}
       \end{equation}
       One can see that the mass $m$ drops out from the equation so that only $c$ and $\beta$ remain parameters.
       In particular, $c$ only multiplies the right-hand side of the BSE and thus its eigenvalue spectrum,
       so that only $t$, $\alpha$ and $\beta$ enter as external parameters.

            \begin{figure}[t]
                    \begin{center}
                    \includegraphics[width=0.7\columnwidth]{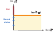}
                    \caption{Singularity structure in the complex $\sqrt{t}$ plane.}\label{fig:sqrt-t-plane}
                    \end{center}
                    \vspace{-5mm}
            \end{figure}

       The homogeneous BSE is an eigenvalue equation
       as it has the formal structure
       \begin{equation}\label{BSE-EV}
         \mathbf{K}\,\mathbf{G_0}\,\Gamma_i = \lambda_i\,\Gamma_i\,,
       \end{equation}
       where the eigenvalues $\lambda_i$ depend on $\sqrt{t}=iM/(2m)$ and the mass ratio $\beta$. Because $\alpha$ is just a momentum partitioning parameter,
       the eigenvalues are independent of $\alpha$ which we also confirmed in our numerical calculations.
       On the other hand, the  dependence of the eigenvalues on $\sqrt{t}$ determines the physical spectrum:
       the BSE has solutions for $\lambda_i(t_i) = 1$, which determine the masses $M_i$ of the ground and excited states.
       Because the coupling strength $c$ in the scalar model is just an overall parameter that is not constrained by anything,
       one can always tune it to produce physical solutions by setting $c=1/\lambda_i(t_i)$. In the following it is therefore not relevant
       \textit{where} a bound state appears; it is sufficient to know that for appropriate values of $c$ they may appear for imaginary $\sqrt{t}$ in the interval
       $0 < \text{Im}\sqrt{t} < 1$, which corresponds to $M<2m$ below the threshold as illustrated in Fig.~\ref{fig:sqrt-t-plane}.

            \begin{figure}[t]
                    \begin{center}
                    \includegraphics[width=0.95\columnwidth]{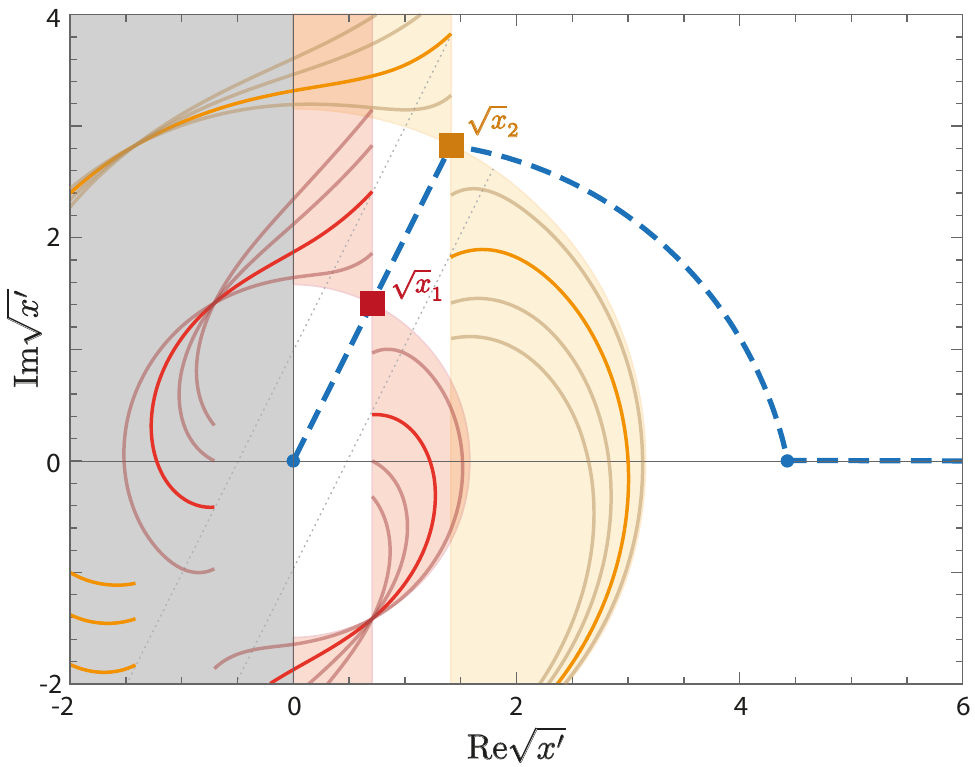}
                    \caption{Branch cuts arising from the Bethe-Salpeter kernel, Eq.~\eqref{kernel-cuts}, in the complex $\sqrt{x'}$ plane.
                    We chose two points $\sqrt{x}_1 = (1+2i)/\sqrt{2}$ and $\sqrt{x}_2 = (1+2i)\sqrt{2}$
                    and four values of $\beta = 0.2, 1, 2, 3$ in each case. The blue dashed line is an integration contour that avoids all possible cuts.}\label{fig:cuts-K}
                    \end{center}
                    \vspace{-5mm}
            \end{figure}

\subsection{Singularities and contour deformations}\label{sec:sing-cd}

             The practical difficulties in solving the BSE, in particular in view of extracting the LFWF, arise from its singularity structure:

             \smallskip

             {\tiny$\blacksquare$} The BSE must be solved for $\sqrt{x}$ inside the  region $\mR_\epsilon$ shown in Fig.~\ref{fig:cuts-7}, because only this region
                 returns the correct result for the LFWF~\eqref{def-lfwf-13} when integrating over real $\omega \in (-\infty,\infty)$.

             \smallskip

             {\tiny$\blacksquare$} The BSE is an integral equation, where the amplitude $\Gamma(x,\omega,t,\alpha)$ is fed back during the iteration.
                  This means it must be solved along a path $\sqrt{x}$ that coincides with the integration path in $\sqrt{x'}$.
                  The natural path is the straight line between $\sqrt{x}=0$ and $\sqrt{x}\to\infty$, but because of the  observations above
                  this path must be deformed into the complex plane to lie entirely within $\mR_\epsilon$.

             \smallskip

            \begin{figure}[t]
                    \begin{center}
                    \includegraphics[width=0.55\columnwidth]{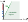}
                    \caption{Calculable region in the complex $\sqrt{t}$ plane when employing contour deformations, for an arbitrary singularity position in the propagator or the amplitude.
                             }\label{fig:iu}
                    \end{center}
                    \vspace{-5mm}
            \end{figure}

             {\tiny$\blacksquare$} The BSE kernel~\eqref{kernel} has a pole at
             \begin{equation}\label{kernel-cuts}
               \sqrt{x'} = \sqrt{x}\left( \Omega \pm i\sqrt{1-\Omega^2 + \frac{\beta}{x}}\right).
             \end{equation}
             After integrating over $\omega'$ and $y$, the pole turns into a branch cut in the complex $\sqrt{x'}$ plane.
             For $\omega, \omega', y \in [-1,1]$, also the variable $\Omega$ is in the interval $\Omega \in [-1,1]$.
             The task of picking the correct residues in analogy to Sec.~\ref{sec:minkowski} then translates into \textit{avoiding} the branch cuts in the  $\sqrt{x'}$ integration.
             The resulting cuts for different values of $\beta>0$ are shown in Fig.~\ref{fig:cuts-K}: For a given point $\sqrt{x}$,
             they are confined to a region bounded by the circle with radius $|\sqrt{x}|$ and a line at $\sqrt{x'} = \text{Re}\sqrt{x}$ (see~\cite{Eichmann:2019dts} for a detailed discussion).
             Because the same happens for \textit{every} point along the integration path (blue dashed line in Fig.~\ref{fig:cuts-K}),
             avoiding all possible cuts for different $\beta$ values implies that once the  path  has reached a particular value $\sqrt{x}_1$, it can only proceed
             if both the real part of $\sqrt{x'}$ and its absolute value do not decrease --- otherwise one would turn back into
             a region populated by branch cuts from the previous point $\sqrt{x}_1$. This limits the possible
             contours in $\sqrt{x}$ on which the BSE is solved: both $\text{Re}\sqrt{x}$ and $|\sqrt{x}|$ must never decrease along such a contour.
             The dashed curve in Fig.~\ref{fig:cuts-K} is an example for a contour satisfying these constraints.

             \smallskip

             {\tiny$\blacksquare$} The singularities arising from the propagators in~\eqref{prop-matrix}
             produce the same cuts $\sqrt{x}_\pm$ as in  Eq.~\eqref{cuts-lfwf-0}, except that the integration variable $\omega$ does not span the full real axis
             but only the interval $\omega \in [-1,1]$. These are already taken care of by choosing a path inside the  region $\mR_\epsilon$.
             In particular, Eqs.~\eqref{cuts-lfwf-0} and~\eqref{kernel-cuts} become identical if one replaces $\Omega\to\omega$, $\beta \to 1$ and $\sqrt{x} \to A_\pm$.
             The resulting cuts have analogous shapes as in Fig.~\ref{fig:cuts-K}, where the two circles  have radii $|1\pm\alpha| \,|\sqrt{t}|$.
             Therefore, a path that ensures safe passage is the one that connects the origin with the point $(1+|\alpha|)\sqrt{t}$
             and then turns back to the real axis by increasing its absolute value as shown in Fig.~\ref{fig:cuts-K}.

             If we are not interested in the LFWF but only the BSWF, then only the interval $\omega \in [-1,1]$ is relevant for the integration.
             From the discussion in Appendix~\ref{app:cuts}, in that case Eq.~\eqref{constraint-1} relaxes to
             \begin{equation}\label{constraint-2}
                \text{Im} \sqrt{t} < \text{min} \left[ \frac{1}{1\pm\alpha}\,, \, \frac{\sqrt{\gamma}}{|\alpha|} \right] \to \text{min} \left[ \frac{1}{2}\,, \, \sqrt{\gamma}\right],
             \end{equation}
             in which case no contour deformations are necessary for any value of $\alpha$. For $\alpha=0$, this reduces to $\text{Im}\sqrt{t} < 1$
             as discussed in~\cite{Eichmann:2019dts}. Note  that the parameter $\gamma$ only applies to the monopole example,
             whereas in the  BSE solution the BS amplitude is calculated dynamically.

            \begin{figure}[t]
                    \begin{center}
                    \includegraphics[width=1\columnwidth]{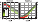}
                    \caption{Largest eigenvalue of the Bethe-Salpeter equation~\eqref{BSE-EV} for $\beta = 4$
                             plotted over $\text{Im}\sqrt{t} = \text{Re}\,M/(2m)$ for six values of $\text{Re}\sqrt{t}$.
                             The physical region for bound states ($0 < M < 2m$) corresponds to $\text{Re}\sqrt{t} = 0$
                             and $0 < \text{Im}\sqrt{t} < 1$.
                             The eigenvalue obtained with the Nakanishi method is shown for comparison
                             for $\text{Re}\sqrt{t} = 0.20$ (red dots).}\label{fig:eigenvalues}
                    \end{center}
                    \vspace{-5mm}
            \end{figure}

             \smallskip

             {\tiny$\blacksquare$} The BS amplitude $\Gamma(x,\omega,t,\alpha)$ may dynamically generate singularities in the course of the iteration.
             The $\omega$ dependence does not cause any trouble because the interval $\omega\in[-1,1]$ is free of singularities.
             However, in principle the equation can generate singularities in the complex $\sqrt{x}$ plane, like the monopole amplitude~\eqref{monopole-E} does
             (the corresponding cut in Fig.~\ref{fig:cuts-7} is the yellow curve for $\sqrt{x}_0$).
             As explained in Appendix~\ref{app:cuts}, a singularity $q^2/m^2 = -u^2$ does not affect the contour deformation if
             \begin{equation}\label{arg-arg}
                \arg(\sqrt{t})  < \arg(iu)\,.
             \end{equation}
             As long as this condition is satisfied,
             a contour deformation connecting the origin $\sqrt{x}=0$ with the point $\sqrt{x}=(1+|\alpha|) \sqrt{t} $
             and returning to the real axis as described above is always possible. For real singularities (imaginary $iu$)
             the calculation is therefore feasible for any $\sqrt{t} \in \mathds{C}$ in the upper right quadrant,
             whereas for complex singularities the calculable domain in $\sqrt{t}$ is restricted to a segment bounded by the first singularity $\sqrt{t} = iu$
             as shown in Fig.~\ref{fig:iu}.
             In other words, one does not \textit{need} to know the actual singularity locations as long as they are restricted to the region~\eqref{arg-arg}.
             The same statement also holds for propagators with complex singularities  which we will analyze in Sec.~\ref{sec:complex-singularities}.

\subsection{Bethe-Salpeter amplitude}

         By implementing the contour deformations described above, the BSE~(\ref{bse-final}--\ref{BSE-EV}) can be solved for any $\sqrt{t} \in \mathds{C}$.
         Fig.~\ref{fig:eigenvalues} shows the (inverse of the) largest eigenvalue $\lambda_0$ corresponding to the ground state.
         As $\text{Re}\sqrt{t}$ becomes smaller, one can see the formation of the branch point at the threshold \mbox{$\sqrt{t} = i$}.
         The ground state is determined by the conditions $\text{Re}\,1/\lambda_0 = c$ and $\text{Im}\,1/\lambda_0 = 0$.
         Depending on the coupling parameter $c$ and the mass ratio $\beta$, it is either a bound state (with $\beta = 4$ in Fig.~\ref{fig:eigenvalues} this happens for $6 \lesssim c \lesssim 11$),
         a tachyon (larger $c$), or a virtual state on the second Riemann sheet (smaller $c$), as shown in~\cite{Eichmann:2019dts} by solving the corresponding scattering equation.

         In the following we  compare our results with those obtained using a Nakanishi representation,
         which has been frequently used in the calculation of light-front quantities~\cite{Nakanishi:1963zz,Nakanishi:1969ph,Nakanishi:1988hp,Kusaka:1995za,Kusaka:1997xd,Sauli:2001we,Karmanov:2005nv,Sauli:2008bn,Carbonell:2010zw,Frederico:2013vga}.
          The central task in that case is the calculation of the Nakanishi weight function
         for a given interaction model, from where all further quantities (BSWF, LFWF, etc.) are obtained;
         see Appendix~\ref{sec:Nakanishi} for details.
         In Fig.~\ref{fig:eigenvalues} one can see that the eigenvalues obtained
         from both methods are in excellent agreement.

            \begin{figure}[t]
                    \begin{center}
                    \includegraphics[width=0.95\columnwidth]{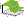}
                    \caption{Absolute value of the BS amplitude for $\alpha = -0.602$ and $\sqrt{t} = 0.20+0.20i$ as a function of $x$ and $\omega$ (all variables are dimensionless).
                     The crosses mark selected values $x_1 = 0.94$, $x_2 = 11$, $x_3 = 101$ and $\omega_1 = 0$, $\omega_2= 0.5$ which are referred to in Fig.~\ref{fig:BSA-alpha-dependence}.}\label{fig:BSA}
                    \end{center}
                    \vspace{-5mm}
            \end{figure}

\begin{figure}[t]
                    \begin{center}
                    \includegraphics[width=1\columnwidth]{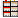}
                    \caption{Dependence of the BS amplitude on $\alpha$ for fixed values of $x$ and $\omega$ as indicated in Fig.~\ref{fig:BSA}.
                             The curves correspond to three different values  $\sqrt{t} = 0.2 + \lambda i$ with $\lambda \in \{0.2, 0.8,1.2\}$.
                             The results for $-\omega$ are identical to those for $+\omega$ when exchanging $\alpha\to-\alpha$.
                              }\label{fig:BSA-alpha-dependence}
                    \end{center}
                    \vspace{-5mm}
            \end{figure}

         The typical shape of the solution for the BS amplitude $\Gamma(x,\omega,t,\alpha)$
         is shown in Fig.~\ref{fig:BSA} for a fixed value of $\alpha$ and $t$.
         The amplitude falls off like a monopole $\propto 1/x$ in the $x$ direction,
         whereas the $\omega$ dependence is very weak and difficult to see in the 3D plot.
         From Eqs.~(\ref{kernel}--\ref{bse-final}) it follows that the amplitude is invariant under a combined  flip $\alpha\to-\alpha$ and $\omega\to-\omega$,
         because together with $\omega' \to -\omega'$ this operation leaves the BSE invariant.
         Also the $\alpha$ dependence is rather modest, as shown in Fig.~\ref{fig:BSA-alpha-dependence} for exemplary values of $x$, $\omega$ and $t$.
         This suggests that the $\alpha$ dependence  of the LFWF~\eqref{def-lfwf-13} is largely carried by the propagator product $G_0$
         entering in the BSWF $\Psi = G_0\,\Gamma$.

         On the other hand, the amplitude can  not be entirely independent of $\omega$  because then the LFWF would no longer vanish at the endpoints  $\alpha =\pm 1$.
         In that case, Eq.~\eqref{def-lfwf-13} reduces to
        \begin{equation}
           \psi(\pm 1,x,t) \propto \int\limits_{-\infty}^\infty d\omega \,\frac{\Gamma(x,\omega,t,\pm 1)}{\omega - \omega_0}  \stackrel{!}{=} 0
        \end{equation}
        with $\omega_0 = \mp (1+x+4t)/(4\sqrt{xt})$. If $\Gamma$ were independent of $\omega$,  this integral would not converge; moreover,
        by Cauchy integration it can only vanish if the amplitude has singularities for some $\omega \in \mathds{C}$ that
        lie on the same (upper or lower) half plane as the pole at $\omega = \omega_0$.
        This requirement coincides with the conditions defining the region $\mR_\epsilon$ in Fig.~\ref{fig:cuts-7} for $\alpha = \pm 1$,
        so the vanishing of the LFWF at the endpoints is automatic as long as $\sqrt{x} \in \mR_\epsilon$.

      \subsection{Light-front wave function}

       We now proceed with the calculation of the LFWF according to Eq.~\eqref{def-lfwf-13}.
       At this point, the  solution for the BS amplitude $\Gamma(x,\omega,t,\alpha)$ and thus
       the BSWF $\Psi(x,\omega,t,\alpha)$  has been determined numerically
       for $\alpha \in [-1,1]$, $\omega \in [-1,1]$ and $\sqrt{x}$ along a contour inside the region $\mR_\epsilon$
       (cf.~Fig.~\ref{fig:cuts-7}). The additional complication is that in the integration to obtain the LFWF one needs to know the dependence on $\omega$ over the whole
             real axis and not just inside the interval $\omega \in [-1,1]$.

            \begin{figure}[p]
                    \begin{center}
                    \includegraphics[width=0.9\columnwidth]{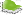}
                    \caption{Absolute value of the light-front wave function as a function of $x$ and $\alpha$, for  $\sqrt{t} = 0.20+0.80i$ and $\beta=4$.}\label{fig:LFWF}
                    \vspace{5mm}
                    \includegraphics[width=1\columnwidth]{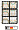}
                    \caption{Dependence of the light-front wave function $\psi(\alpha,x,t)$ on $\alpha$ for three different values $x = 0.039$ (top), $x=0.326$ (center) and $x=2.05$ (bottom)
                             and different values of $\sqrt{t}$. The results obtained with contour deformations and the SPM
                             are compared to the Nakanishi results; the latter are restricted to $\text{Im}\sqrt{t} < 1$ below the threshold.}\label{fig:LFWF-2}
                    \end{center}
                    \vspace{-5mm}
            \end{figure}

          To analytically continue the $\omega$ dependence to the entire real axis,
          we employ the Schlessinger point method (SPM)~\cite{Schlessinger:1968}
          which has found widespread recent applications as a high-quality tool for analytic continuations
          into the complex plane, see e.g.~\cite{Haritan:2017vvv,Tripolt:2017pzb,Tripolt:2018xeo,Eichmann:2019dts,Santowsky:2020pwd,Huber:2020ngt,Cui:2021vgm,Cui:2021gzg,Huber:2021yfy}. It amounts to a continued fraction
          \begin{equation}\label{SPM}
            f(\omega) = \frac{c_1 \qquad\qquad}{1 + \displaystyle\frac{c_2\,(\omega-\omega_1)}{1 + \displaystyle\frac{c_3\,(\omega-\omega_2)}{1 + \displaystyle\frac{c_4\,(\omega-\omega_3)}{\dots}}}}\,,
          \end{equation}
          which is simple to implement by an iterative algorithm.
          Given $n$ input points $\omega_i$ with $i=1 \dots n$ and a function whose values $f(\omega_i)$ are known,
          one determines the $n$ coefficients $c_i$ and thereby obtains an analytic continuation
          of the original function for arbitrary values $\omega\in\mathds{C}$.
          The continued fraction can be recast into a standard Pad\'e form in terms of a division of two polynomials.

          In our case, $f(\omega)$ is the BS amplitude $\Gamma(x,\omega,t,\alpha)$ for fixed values of $x$, $t$ and $\alpha$.
          The input points $\omega_i$ lie inside the interval $\omega_i \in [-1,1]$,
          and the analytic continuation is performed for $\omega \in \mathds{R}$.
          The LFWF~\eqref{def-lfwf-13} is finally obtained by integrating the resulting BSWF over $\omega \in \mathds{R}$.

            \begin{figure}[t]
                    \begin{center}
                    \includegraphics[width=1\columnwidth]{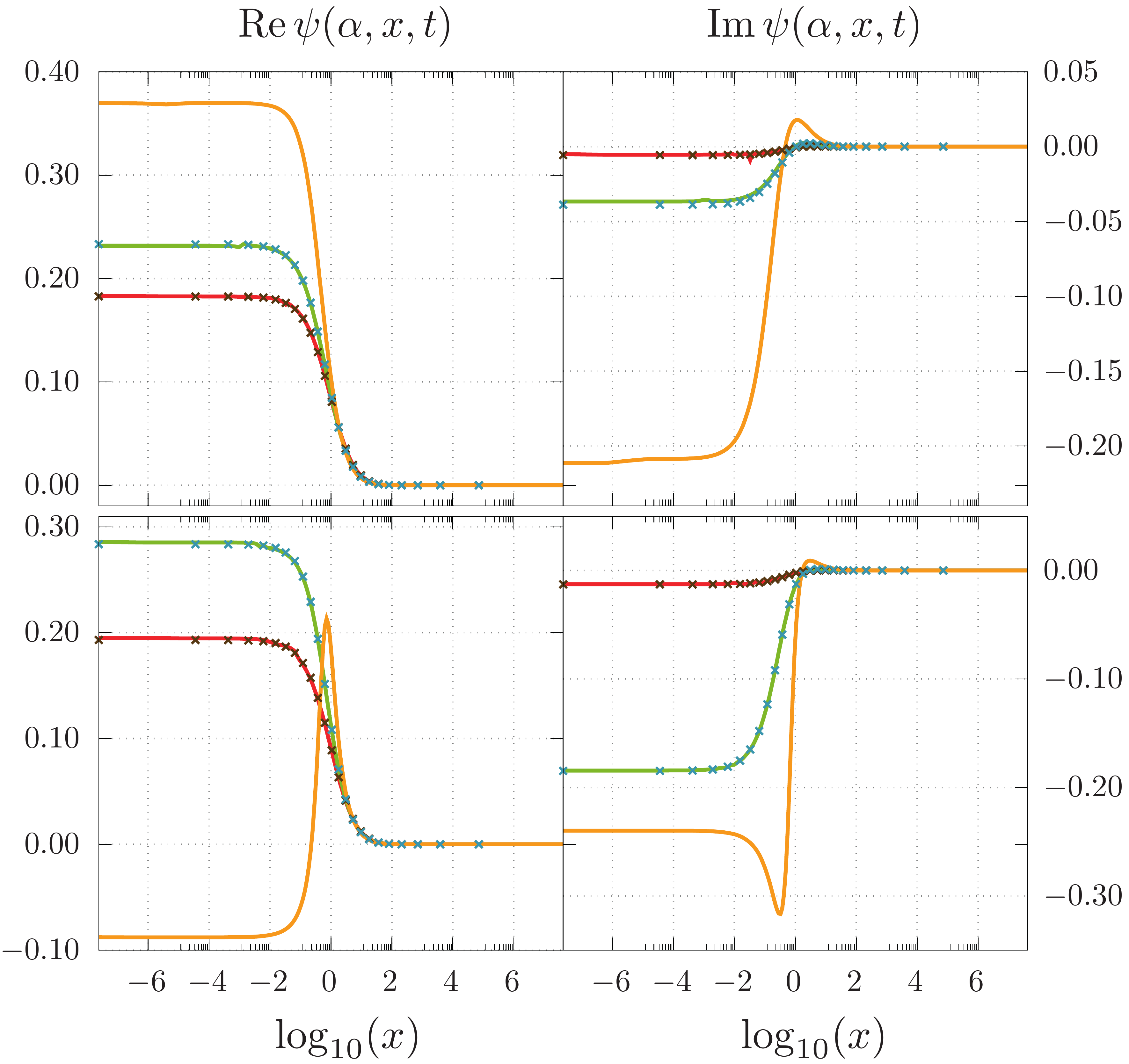}
                    \caption{Same as in Fig.~\ref{fig:LFWF-2}, but now as a function of $x$ for fixed $\alpha = 0.668$ (top) and $\alpha = 0.043$ (bottom).}\label{fig:LFWF-3}
                    \end{center}
                    \vspace{-5mm}
            \end{figure}

             The resulting LFWF for an exemplary value of $\sqrt{t}$ is shown in Fig.~\ref{fig:LFWF}.
             The falloff in $x$, the symmetry in $\alpha$ and the vanishing at the endpoints $\alpha = \pm 1$ are clearly visible.
             We did not employ any polynomial expansion to facilitate the calculation; the result is the plain integral from Eq.~\eqref{def-lfwf-13}.
             In Fig.~\ref{fig:LFWF} and also the subsequent plots we employed an additional SPM step analogous to Eq.~\eqref{SPM}
             to transform the LFWF, which is obtained along a complex contour in $x$, to the real axis $x \in \mathds{R}_+$.

             Figs.~\ref{fig:LFWF-2} and ~\ref{fig:LFWF-3} give a detailed view of the $\alpha$ dependence of the LFWF (for selected values of $x$)
             and its $x$ dependence (for selected values of $\alpha$), respectively.
             The curves correspond to three values of $\sqrt{t} \in \mathds{C}$, two of which lie below the threshold ($\text{Im}\sqrt{t} = 1$) and one above.
             The results  agree very well with the Nakanishi method which is applicable below the threshold.
             The region above the threshold is unphysical since there are no physical poles on the first sheet
             (the scalar model also does not produce resonances but instead virtual states on the second sheet~\cite{Eichmann:2019dts}), but
             one can see in the plots that the LFWF is well-defined and calculable for any value of $\sqrt{t}$.

          A necessary condition for our strategy to be applicable is  that the BS amplitude does not have singularities for $\omega \in \mathds{R}$.
          We  checked that this is indeed not the case.
          Although the SPM cannot produce branch cuts but only poles, the resulting pole structure indicates the existence of  cuts
          connecting the origin with the point $\omega = \sqrt{t}$ for $|\omega|>1$.
             However, we note that even if the BS amplitude \textit{had} singularities for real $\omega$,
             this would not invalidate the approach because one would merely need to rotate the integration contour in Eq.~\eqref{def-lfwf-13} accordingly.
             In that case the region $\mR_\epsilon$ in Fig.~\ref{fig:cuts-7} would also change and one would need to solve the BSE
             along a modified path in $\sqrt{x}$.

            \begin{figure}[t]
                    \begin{center}
                    \includegraphics[width=1\columnwidth]{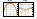}
                    \caption{Distribution amplitude $\phi(\alpha)$ from contour deformations (solid lines) compared to the results from the Nakanishi method (crosses);
                             see Fig.~\ref{fig:LFWF-2} for the legend.}\label{fig:pda}
                    \end{center}
                    \vspace{-5mm}
            \end{figure}

            We close this section with some remarks on the numerical stability.
            The two relevant variables in the BSE~\eqref{bse-final} are the radial variable $\sqrt{x}$,
            which takes values along the deformed contour shown in Fig.~\ref{fig:cuts-K}, and the angular variable $\omega \in [-1,1]$.
            For each of the three segments in $\sqrt{x}$ we employed a Gauss-Legendre quadrature % that maps each segment to the interval $[-1,1]$,
            with $N_x$ integration points in total.
            For the $\omega$ direction we employed a Chebyshev quadrature with $N_\omega$ points.
%            The number of points $N_x$ and $N_\omega$ to reach sub-permille precision
            The necessary number of integration points depends on the external variables $\sqrt{t}$ and $\beta$,
            where $\beta$ is the mass ratio from Eq.~\eqref{c-beta}.
            If $\sqrt{t}$ comes close to the imaginary axis, one needs more integration points
            since the poles in the integrand and resulting cuts move closer to the imaginary axis and the integration path (left panel in Fig.~\ref{fig:cuts-7}).
            Similarly, for small values of $\beta$ the cuts from the kernel in Fig.~\ref{fig:cuts-K} become increasingly circular
            and for $\beta=0$ the points along the integration path itself become the branch points.
            For not too small values of $\text{Re}\sqrt{t} \gtrsim 0.1$ and $\beta \gtrsim 1$,
            we find that $N_x = 96$ and $N_\omega = 64$ is sufficient to reach sub-permille precision for the BSE eigenvalues in Fig.~\ref{fig:eigenvalues}.
            For smaller $\sqrt{t}$ or $\beta$ these numbers increase, and the numerical stability is typically more
            sensitive to $N_x$ than $N_\omega$. The same values
            are sufficient to achieve numerical convergence for the LFWF, which is also more sensitive to $N_x$ than $N_\omega$.
            Finally, for the analytic continuation in $\omega$ using the SPM  % that we used in the analytic continuation from $\omega\in[-1,1]$ to $\omega\in\mathds{R}$,
            we found an optimal value $N_\text{SPM} = 24$ for the number of input points to achieve agreement with the Nakanishi method.

%            Concerning the SPM~\eqref{SPM} that we used in the analytic continuation from $\omega\in[-1,1]$ to $\omega\in\mathds{R}$,
%            we found an optimal value $N_\text{SPM} = 24$ for the number of input points to achieve agreement with the Nakanishi method.

            \begin{figure}[t]
                    \begin{center}
                    \includegraphics[width=0.9\columnwidth]{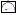}
                    \caption{Distribution amplitude $\phi(\alpha)$ for $\sqrt{t} = 0.5i$ in the physical region.
                             The band is the result using contour deformations and analytic continuations and compared to the result  from the Nakanishi method.}\label{fig:pda-phys}
                    \end{center}
                    \vspace{-0mm}
            \end{figure}

\subsection{Light-front distributions}

            The remaining task is to compute the distribution amplitude $\phi(\alpha)$ and distribution function $u(\alpha)$ as given in Eq.~\eqref{da-df-2}.
            This amounts to an integration of the LFWF over the transverse momentum variable $x$, which does not need to lie on the real axis
            since one can integrate along the same deformed contour on which the LFWF was obtained.

            The resulting distribution amplitude $\phi(\alpha)$ is shown in Fig.~\ref{fig:pda},
            again for three values of $\sqrt{t} \in \mathds{C}$ and compared to the results obtained with the Nakanishi method.
            It inherits the same properties as the  LFWF; once again, the figure shows the plain numerical result
            without any expansion in moments.
            Also here the contour deformation method is in good agreement with the results from the Nakanishi method.

            Although we did not employ them in our calculations, one usually defines Mellin moments $\langle \xi^m \rangle$
            for the reconstruction of hadronic distribution functions through the momentum fraction $\xi$, cf.~Eq.~\eqref{xi}:
            \begin{equation}
            \begin{split}
                \langle \xi^m \rangle &= \int_0^1 d\xi \,\xi^m \,\phi(\alpha) = \frac{1}{2}\int_{-1}^1 d\alpha\left( \frac{1+\alpha}{2}\right)^m \phi(\alpha) \\
                                      &= \frac{1}{2^{m+1}}\sum_{k=0}^m \left(m \atop k\right) \int_{-1}^1 d\alpha\,\alpha^k\,\phi(\alpha)\,.
            \end{split}
            \end{equation}
            For the plots we used the normalization $\int d\alpha\,\phi(\alpha) = 2$, which ensures $\langle \xi^0 \rangle = 1$ for the zeroth  moment.
            The first moment $\langle \xi\rangle = 1/2$ is  the mean value of the distribution which is centered around $\xi=1/2$ or $\alpha = 0$.

   \pagebreak

            \begin{figure}[t]
                    \begin{center}
                    \includegraphics[width=0.85\columnwidth]{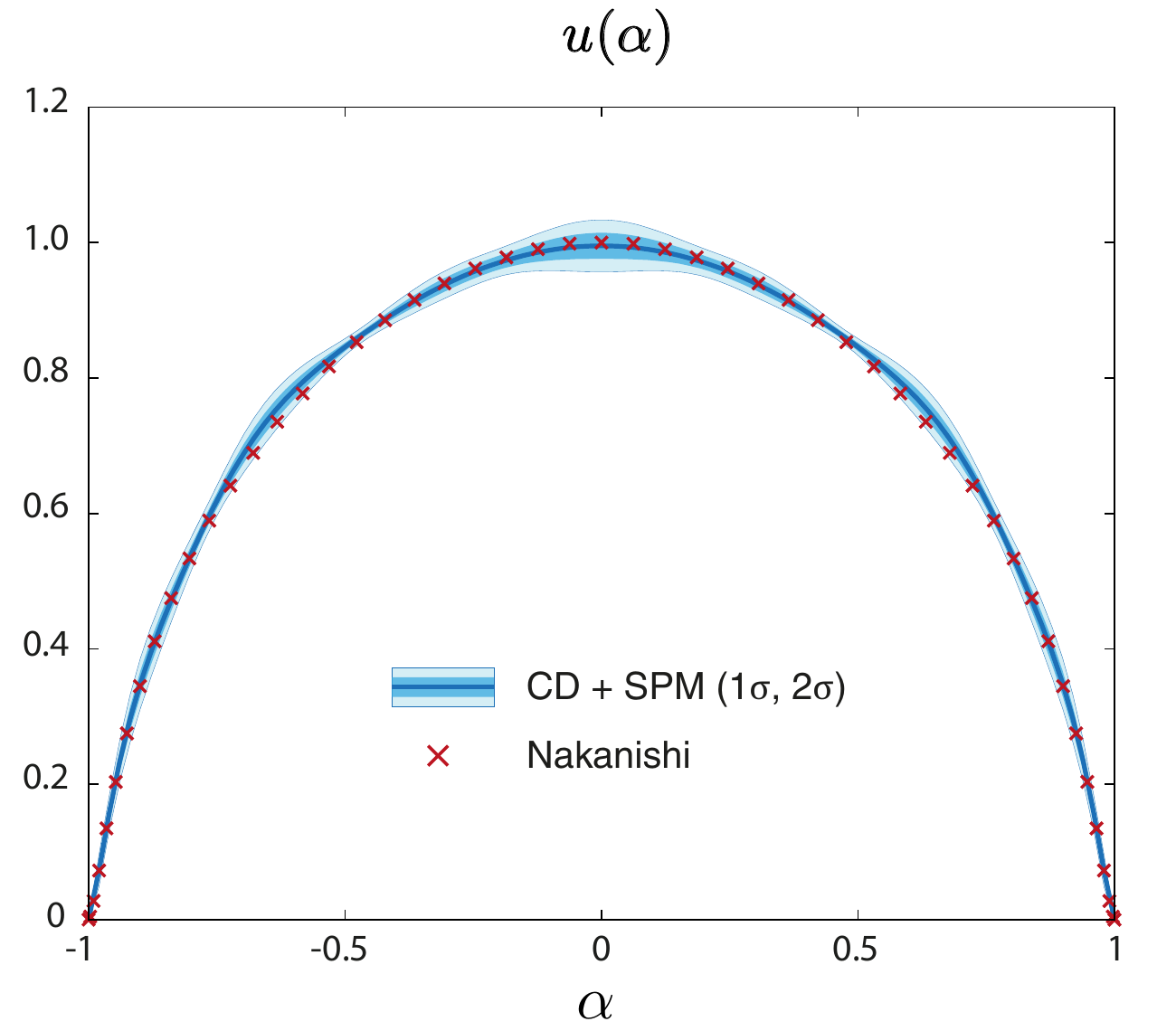}
                    \caption{Distribution function $u(\alpha)$ for $\sqrt{t} = 0.5i$ in the physical region, normalized to $u(0)=1$.}\label{fig:pdf-phys}
                    \end{center}
                    \vspace{-5mm}
            \end{figure}

            All results presented so far have been obtained for $\sqrt{t}\in \mathds{C}$ in the upper right quadrant.
            As discussed above in connection with Figs.~\ref{fig:cuts-7} and~\ref{fig:sqrt-t-plane}, it is numerically not possible to calculate the LFWF and PDA
            directly for imaginary $\sqrt{t}$ corresponding to real masses $0 < M < 2m$,
            because in that case the only allowed integration path would coincide with the branch cuts.

            To compute the PDA in the physical region, we employ the SPM from Eq.~\eqref{SPM}
            to analytically continue the results for complex $\sqrt{t}$ to the imaginary axis in $\sqrt{t}$.
            Here we employed a Chebyshev expansion
            \begin{equation}
                \phi(\alpha) = (1-\alpha^2) \sum_n \phi_n\,U_n(\alpha)\,,
            \end{equation}
            where $U_n(\alpha)$ are the Chebyshev polynomials of the second kind,
            and performed the analytic continuation for the Chebyshev moments $\phi_n$. % to ensure numerical stability.
            We chose an  input region $\text{Re}\sqrt{t} \geq 0.1$ for the SPM and changed the number of input points from $N= 10$ to $N=50$ in steps of $\Delta N=2$.
            The resulting PDA at $\sqrt{t} = 0.5i$ is shown in Fig.~\ref{fig:pda-phys}, where the mean value is the average over the different input points and
            the error bands are the $1\sigma$ and $2\sigma$ deviations.
            Also here the results match with the Nakanishi method.
            The analogous result for the distribution function $u(\alpha)$ is shown in Fig.~\ref{fig:pdf-phys} and displays a somewhat larger error
            from the analytic continuation.

            The SPM is reliable for imaginary values of $\sqrt{t}$ sufficiently below the threshold, whereas
            above the threshold it would attempt to continue to the second Riemann sheet and thus the results deteriorate as one moves closer to the threshold.
            We emphasize, however, that an analytic continuation in $\sqrt{t}$
            is not necessary in principle since all quantities of interest can be calculated directly but at the expense of an increasing numerical cost for $\text{Re}\sqrt{t} \to 0$.
            In any case, for the scope of this exploratory study it is clear that the contour-deformation method  %from Figs.~\ref{fig:pda}--\ref{fig:pdf-phys}
            is well suited to compute light-front distributions in a quantitatively reliable manner.

    \newpage

            \begin{figure}[t]
                    \begin{center}
                    \includegraphics[width=1\columnwidth]{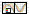}
                    \caption{Light-front wave function for unequal masses and five different values of $\varepsilon$,
                             evaluated at $x=1.22$, $\sqrt{t} = 0.2+0.8i$ and $\beta=4$.}\label{fig:LFWF-unequal}
                    \end{center}
                    \vspace{-5mm}
            \end{figure}

      \section{Generalizations}\label{sec:generalizations}

      In this section we explore two extensions of the contour-deformation method within the scalar model
      to bridge the gap towards possible future applications in QCD: One is the generalization to unequal masses in the BSE
      and the other is the implementation of complex conjugate propagator singularities.

      \subsection{Unequal masses}

      A straightforward generalization is the case of unequal masses $m_1 \neq m_2$ of the constituents in the two-body BSE.
      To this end we write
      \begin{equation}
         m_1 = m\,(1+\varepsilon)\,, \quad m_2 = m\,(1-\varepsilon)
      \end{equation}
      such that
      \begin{equation}
        m = \frac{m_1 + m_2}{2}\,, \quad \varepsilon = \frac{m_1-m_2}{m_1+m_2}\,.
      \end{equation}
      In this way the mass parameter $m$  drops out from the BSE like before, which in turn depends on $\varepsilon$.

      Alternatively, one may start from Eq.~\eqref{momenta-11} with the original momenta $q$ and $P$ (instead of $k$ and $P$)
      with an arbitrary momentum partitioning parameter $\varepsilon \in [-1,1]$. For unequal masses, the choice $\varepsilon = (m_1-m_2)/(m_1+m_2)$ maximizes
      the domain in $t$ where the BSE can be solved without contour deformations, namely $\text{Im}\sqrt{t} < 1$.
      When introducing the momentum $k$ through Eqs.~(\ref{q-k-splitting}--\ref{momenta-12}),
      the momentum partioning parameter $\varepsilon$ drops out from all subsequent equations in favor of $\alpha$ which takes its place,
      and setting $q^+ = (\alpha-\varepsilon) P^+/2$ in the LFWF leads to the same result as before, Eq.~\eqref{def-lfwf-13}.
      Thus, $\varepsilon$ only appears in the propagator product
       \begin{equation}
            G_0(q,P) = \frac{1}{q_1^2+m_1^2}\,\frac{1}{q_2^2+m_2^2}
      \end{equation}
      through the masses $m_1$ and $m_2$ and assumes the role of the physical mass difference as defined above.

  \begin{figure*}[t]
  \includegraphics[width=0.85\textwidth]{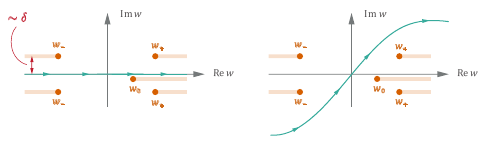}
  \caption{Complex conjugate propagator poles in the complex $w$ plane. The interpretation of the $i\epsilon$ prescription like in Fig.~\ref{fig-int-path}(a) does not return
           the correct limit $\delta\to 0$ (left), in contrast to the one according to Fig.~\ref{fig-int-path}(b) (right).}
  \label{fig-complex-poles}
  \end{figure*}

      Writing $u_\pm = 1 \pm \varepsilon$,
      Eqs.~(\ref{kinematics-E2}--\ref{cuts-lfwf-0}) generalize to
         \begin{equation}
         \begin{split}
            \frac{q_{1,2}^2}{m^2} +u_\pm^2 &= x - 2\sqrt{x}\, \omega A_\pm + A_\pm^2  + u_\pm^2 \,, \\
            \sqrt{x}_\pm &= A_\pm\left[ \omega + i\lambda \sqrt{1-\omega^2 + \frac{u_\pm^2}{A_\pm^2}}\,\right] ,
         \end{split}
         \end{equation}
      with $A_\pm  = \mp (1\pm \alpha)\sqrt{t}$ as before.
      This does not require any modifications in the contour deformation:
      a path connecting the origin with the point $(1+|\alpha|)\sqrt{t}$
      and turning back to the real axis by increasing its real part and absolute value is  sufficient to avoid the cuts for any $u_\pm \in \mathds{R}$,
      so it is equally applicable in this case.

      In the unequal-mass case, the BSE eigenvalues are still independent of $\alpha$ but they change with $\varepsilon$,
      which is a physical parameter in the system. The BS amplitude is still invariant under the combined flip
      $\alpha\to-\alpha$, $\omega\to-\omega$ and $\varepsilon\to-\varepsilon$.
      Fig.~\ref{fig:LFWF-unequal} shows the resulting LFWF for different values of $\varepsilon$ and selected
      values of $x$ and $t$. In the unequal-mass case, the LWFW and corresponding light-front distributions are
      no longer centered at $\alpha=0$ but tilted towards the heavier particle.

      \subsection{Complex propagator singularities}\label{sec:complex-singularities}

       Another generalization concerns complex singularities in the propagators,
       which is the typical situation for QCD propagators obtained from functional methods within truncations, see e.g.~\cite{Maris:1997tm,Alkofer:2003jj,Eichmann:2016yit,Windisch:2016iud,Fischer:2020xnb}.
       For a single complex pole pair the propagator becomes
       \begin{align}
           D(q^2) &= \frac{1}{2}\left( \frac{1}{q^2 + m^2\,(1+i\delta)} + \frac{1}{q^2+m^2\,(1-i\delta)}\right) \nonumber \\
                  &= \frac{q^2+m^2}{(q^2+m^2)^2 + m^4\,\delta^2} \,, \qquad \delta > 0 \,,
       \end{align}
      which for $\delta = 0$  reduces to a  real mass pole.
      Writing $u^2 = 1 + i \bar{\lambda}\delta$ with $\bar\lambda = \pm 1$,
      Eqs.~(\ref{kinematics-E2}--\ref{cuts-lfwf-0}) generalize to
         \begin{equation}
         \begin{split}
            \frac{q_{1,2}^2}{m^2} +u^2 &= x - 2\sqrt{x}\, \omega A_\pm + A_\pm^2  + u^2 \,, \\
            \sqrt{x}_\pm &= A_\pm\left[ \omega + i\lambda \sqrt{1-\omega^2 + \frac{u^2}{A_\pm^2}}\,\right] ,
         \end{split}
         \end{equation}
         again with $A_\pm  = \mp (1\pm \alpha)\sqrt{t}$.
         This can be generalized to the case discussed in Appendix~\ref{app:cuts}, cf.~Fig.~\ref{fig:cuts-5}: For any singularity at $q^2/m^2 = -u^2 \in \mathds{C}$ in the upper half plane, $iu$ lies in the right upper quadrant.
      A contour connecting the origin with the point $(1+|\alpha|)\sqrt{t}$
      and turning back to the real axis by increasing its real and absolute value is  sufficient to avoid the cuts as long as $\arg(\sqrt{t})  < \arg(iu)$ like in Fig.~\ref{fig:iu}.
      If this condition is not satisfied, the cuts twirl in the opposite direction and different cuts may overlap
      so that no contour can be found, i.e., the region $\mR_\epsilon$ no longer connects the origin with infinity.
      Therefore, the appearance of complex singularities does not impede the contour deformation method
      but merely restricts the calculable domain in the complex $\sqrt{t}$ plane.

      Some care needs to be taken, however, when working with complex conjugate poles in Minkowski space using residue calculus like in Sec.~\ref{sec:lfwf-monopole} (see also Ref.~\cite{Tiburzi:2003ja} for a  discussion).
      In that case the literal $i\epsilon$ prescription, where the $i\epsilon$ factors appear in the denominators and one integrates over the real $w$ axis, is no longer meaningful.
      The $w_\pm$ poles in the complex $w$ plane are now separated from the real axis by a finite distance proportional to $\delta$,
      \begin{equation}
         w_\pm^{\bar\lambda} = \pm \left( t + \frac{x+1+i\bar{\lambda}\delta-i\epsilon}{1\pm\alpha}\right),
      \end{equation}
      and the infinitesimal $\epsilon$ term does not change that. An integration over the real axis as shown in the left of Fig.~\ref{fig-complex-poles}
       therefore does not give the correct limit for $\delta \to 0$. By contrast, the integration path in the right panel
      corresponding to the $i\epsilon$ prescription in Eq.~\eqref{lfwf-3a} yields a proper analytic continuation
      and returns the correct limit $\delta \to 0$. This is, of course, equivalent to  the Euclidean integration path
      as long as one stays in the region $\mR_\epsilon$.

\begin{figure}[t]
                    \begin{center}
                    \includegraphics[width=1\columnwidth]{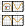}
                    \caption{Light-front wave function (top) and distribution amplitude (bottom) for a propagator with a complex conjugate pole pair.
                             The results correspond to $\sqrt{t} = 0.2+0.8i$ and $\beta=4$; the light-front wave function is evaluated at $x=0.95$.
                             }\label{fig:LFWF-complex}
                    \end{center}
                    \vspace{-7mm}
            \end{figure}

      Fig.~\ref{fig:LFWF-complex} shows the resulting LFWF and PDA obtained from the BSE with a complex conjugate pole pair in the propagators
      and for three values of $\delta$. The results are very similar, which confirms that the LFWF and the quantities derived from it are insensitive to whether the propagator poles are real or complex.
      In conclusion, one can tackle complex singularities just like real singularities
      using contour deformations.

      \section{Summary and outlook}\label{sec:summary}

      In this work we explored a new method to compute light-front quantities
      based on contour deformations and analytic continuations. We applied the method to calculate the light-front wave functions and distributions
      for a scalar model, whose Bethe-Salpeter equation we solved dynamically, and  found excellent agreement with the well-established Nakanishi method.

      The method is quite efficient and  has several advantages,
      as it is not restricted to  bound-state masses below the threshold and
      it can also handle complex singularities in the integrands.
      Although we exemplified the technique for a situation where the propagators are known explicitly,
      in principle one does not even need to know the singularity locations as long as
      they are restricted to a certain kinematical region.
      Since the method is independent of the type of correlation functions,
      in principle one can extend it to the calculation of parton distributions
      such as PDFs, TMDs and GPDs in the future.

      \medskip

\textbf{Acknowledgments.}
We are grateful to Tobias Frederico for valuable comments.
This work was supported by the FCT Investigator Grant IF/00898/2015 and
the Advance Computing Grant CPCA/A0/7291/2020.

  \appendix

  \section{Nakanishi representation}\label{sec:Nakanishi}

  In this appendix we collect the relevant formulas for the Nakanishi method, which provides an alternative way for solving the BSE based on a generalized spectral
  representation~\cite{Nakanishi:1963zz,Nakanishi:1969ph,Nakanishi:1988hp,Kusaka:1995za,Kusaka:1997xd,Sauli:2001we,Karmanov:2005nv,Sauli:2008bn,Carbonell:2010zw,Frederico:2011ws,Frederico:2013vga,Gutierrez:2016ixt,dePaula:2016oct,dePaula:2017ikc,AlvarengaNogueira:2019zcs}.
  The main idea is to express the BSWF through a non-singular Nakanishi weight function $g(x,\alpha)$, multiplied with a denominator that absorbs the analytic structure.
  In Euclidean conventions this amounts to
  \begin{equation}
     \Psi(q,P) = \frac{1}{m^4} \int_0^\infty dx' \int_{-1}^1 d\alpha'\,\frac{g(x',\alpha')}{\left[ \kappa + B(x',\alpha') \right]^3} \,,
  \end{equation}
  where $\kappa = (q-\alpha' P/2)^2/m^2$, $B(x,\alpha) = 1+x+(1-\alpha^2)\,t$ and $t=P^2/(4m^2) = -M^2/(4m^2)$.
  The desired light-front quantities are then obtained through integrations over the weight function;
  e.g., the LFWF~\eqref{def-lfwf-3} becomes
  \begin{equation}
     \psi(x,\alpha) = \frac{\mN}{m^2}\int_0^\infty dx' \frac{g(x',\alpha)}{\left[ x' + B(x,\alpha)\right]^2}\,.
  \end{equation}
  The BSE~\eqref{bse-general} can be rewritten as a self-consistent equation for the weight function $g(x,\alpha)$,
  \begin{equation}\label{bse-nakanishi}
  \begin{split}
     &\int_0^\infty dx' \frac{g(x',\alpha)}{\left[ x' + B(x,\alpha)\right]^2} \\
     & \qquad = c\int_0^\infty dx'\int_{-1}^1 d\alpha'\,V(x,x',\alpha,\alpha')\,g(x',\alpha')\,,
  \end{split}
  \end{equation}
  where $c$ is the global coupling parameter from Eq.~\eqref{c-beta} and the kernel $V$ for a scalar one-boson exchange takes the form~\cite{Karmanov:2005nv,Frederico:2013vga}
  \begin{align}
     & V(x,x',\alpha,\alpha') = \frac{K(x,x',\alpha,\alpha') + K(x,x',-\alpha,-\alpha')}{2B(x,\alpha)}\,, \nonumber \\
     & K(x,x',\alpha,\alpha') = \int_0^1 dv\,\frac{\theta(\alpha-\alpha')\,(1-\alpha)^2}{\left[v(1-\alpha) \,B(x',\alpha')+(1-v)\,C\right]^2}\,, \nonumber \\
     & C =(1-\alpha')\, B(x,\alpha) + (1-\alpha)\left( \frac{\beta}{v}+x'\right).
  \end{align}
  Eq.~\eqref{bse-nakanishi} has the structure of a generalized eigenvalue problem for the weight function $g$,
  \begin{equation}
     \lambda_0\,\mathbf{B} \,g = \mathbf{K} \,g \quad \Rightarrow \quad \mathbf{B}^{-1} \mathbf{K} \,g = \lambda_0 \,g\,,
  \end{equation}
  which can be solved in analogy to the standard BSE except for the additional matrix inversion.

            \begin{figure*}[t]
                    \begin{center}
                    \includegraphics[width=0.90\textwidth]{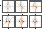}
                    \caption{Shape of the function~\eqref{cut-function} for $A=i$ and different values of $C = 0.6\,\exp\,( i\sigma \frac{\pi}{2})$;
                             from top left to bottom right: $\sigma = 0$, $0.3$, $0.5$, $0.7$, $0.9$, $1$. The blue (red) curves are the
                             positive (negative) branches. The open points are the values of $\pm AC$, the orange circles have radii $|AC|$ and the gray circles radii $|A|$, and the orange lines
                             connect the origin with the points $AC^2$. The filled squares show the intersections with the real axis.}\label{fig:cuts-4}
                    \end{center}
                    \vspace{-5mm}
            \end{figure*}

   \newpage

       \section{Cuts}\label{app:cuts}

       In this appendix we analyze the branch cuts in Eq.~\eqref{cuts-lfwf-0} in more detail.
       Their general form is
       \begin{equation}\label{cut-function}
          \sqrt{x}_\pm = A\left[ \omega \pm \sqrt{\omega^2 + C^2}\right],
       \end{equation}
       where $A = a+ib$ and $C=c+id$ are complex numbers ($a, b, c, d \in \mathds{R}$) and $\omega \in \mathds{R}$ is varied over the real axis.
       This defines a curve in the complex plane whose shape depends on  $A$ and $C$.
       Because $\sqrt{x}_\pm$ is the solution of the equation
       \begin{equation}\label{cut-eq}
          x - 2\sqrt{x}\, \omega A - A^2 \, C^2 = 0\,,
       \end{equation}
       solving for $\omega$ and setting $\text{Im}\,\omega = 0$ gives an alternative parametrization of the cuts which separate
       the two regions in $\sqrt{x}$:
       \begin{equation}\label{cut-regions}
       |x|\,\text{Im}\,(\sqrt{x} A^\ast) = |A|^2 \,\text{Im}\,(\sqrt{x^\ast} A C^2 )\,.
       \end{equation}
       Because only $C^2$ enters in the formulas,
       we restrict $C$ to the upper right quadrant, i.e., $c>0$ and $d>0$. For $C$ in the lower right quadrant the whole structure is mirrored
       along the line $\sqrt{x} = \lambda A$ with $\lambda \in \mathds{R}$.

       In the six panels of Fig.~\ref{fig:cuts-4},
       $|C|$ is fixed and the phase of $C$ varies from $0$ to $\pi/2$
       (in the figure we set $A=i$).
       To facilitate the construction, the open points indicate
        the values $\sqrt{x} = \pm AC$, and
       we draw the circles with radius $|AC|$ and  lines passing through the points $AC^2$ (both in orange).
       The shape of $\sqrt{x}_\pm$ is then as follows:

       \smallskip

       {\tiny$\blacksquare$} The line connecting the origin with $\sqrt{x} = A$
       (in Fig.~\ref{fig:cuts-4}, the imaginary axis) separates two half planes, where the positive branch $\sqrt{x}_+$ is always confined to one side
       and the negative branch $\sqrt{x}_-$ to the other side.

       \smallskip

       {\tiny$\blacksquare$} The positive branch (drawn in blue) starts at the origin $\sqrt{x}_+=0$ (corresponding to $\omega \to -\infty$),
       passes through the point $\sqrt{x}_+=AC$ at $\omega=0$ (open blue points) and goes to $\sqrt{x}_+ \to A\infty$ for $\omega\to\infty$.

       \smallskip

       {\tiny$\blacksquare$} The negative branch (red) starts at $\sqrt{x}_- \to -A\infty$ (for $\omega\to-\infty$), passes through the point $\sqrt{x}_-=-AC$ at $\omega=0$ (open red points) and ends at the origin ($\omega\to\infty$).

       \smallskip

       {\tiny$\blacksquare$} Near the origin, the two branches follow the line in the direction $\sqrt{x} =AC^2$.
       For $C \in \mathds{R}_+$, this is the direction of $A$ (imaginary axis in Fig.~\ref{fig:cuts-4}, top left panel).
       When we rotate $C$ into the complex plane with $\arg C > 0$, the line also rotates and drags the curves with it.

       \smallskip

       {\tiny$\blacksquare$} As $C$ is rotated further, then  for $\text{Im}\,(AC^2) < 0$ the curve $\sqrt{x}_\pm$ eventually crosses the real axis (filled squares in Fig.~\ref{fig:cuts-4}).
       For $A=i$  this means $\text{Re}\,C^2 = c^2 - d^2 < 0$, i.e., $\arg C > \pi/4$.
       For general $A$ the crossing happens at
       \begin{equation}\label{curve-crosses-zero}
       \begin{split}
          |\sqrt{x}_\pm| &=  |A|\,\sqrt{d^2-c^2-2cdr} \,, \\
          |\omega| &=  \frac{|b\,(c+dr)(d-cr)|}{|\sqrt{x}_\pm|}
       \end{split}
       \end{equation}
       with $r=a/b$.
             In a situation where one needs to integrate $\sqrt{x}$ from zero to infinity, a contour deformation is thus necessary.
             The angular region between the lines $\lambda A$ and $-\lambda AC^2$ ($\lambda>0$) is guaranteed to be free of any cuts and thus
             an integration path leading away from the origin in that direction is safe.

       \smallskip

       {\tiny$\blacksquare$} Finally, if $C$ becomes imaginary ($C=id$), the orange line has rotated by $\pi$ and is again the direction of $A$
             (bottom right panel in Fig.~\ref{fig:cuts-4}).
             For $|\omega| < d$, the curves $\sqrt{x}_\pm$ lie on the circle  with radius $|AC|$; for $|\omega| > d$ they follow the line in the direction of $A$.

       \smallskip

\begin{figure}[t]
                    \begin{center}
                    \includegraphics[width=1\columnwidth]{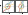}
                    \caption{Shape of the function~\eqref{cut-function-2} for $iu = 0.3+0.6i$ and two values of $A$.
                    %with $\arg{(A)}$ smaller or larger than $\arg{(iu)}$.
                    Depending on $\arg{(A)} \lessgtr \arg{(iu)}$, the branch cuts twirl in different directions.
                    }\label{fig:cuts-5}
                    \end{center}
                    \vspace{-5mm}
            \end{figure}

       The cuts in Eq.~\eqref{cuts-lfwf-0} correspond to
       $C^2 = -1 - u^2/A^2$, where $u^2=1$ for the propagator cuts and $u^2 = \gamma$ for the monopole cut,
       and $A \propto \sqrt{t}$ up to real constants.
       Eq.~\eqref{cut-function} then turns into
       \begin{equation}\label{cut-function-2}
          \sqrt{x}_\pm = A\left[ \omega \pm i\sqrt{1-\omega^2 + \frac{u^2}{A^2}}\,\right] ,
       \end{equation}
       and Eqs.~(\ref{cut-eq}--\ref{cut-regions}) become
       \begin{equation}
       \begin{split}
          x - 2\sqrt{x}\, \omega A + A^2  + u^2 &= 0\,, \\
          \left( |x| - |A|^2 \right) \text{Im}\,(\sqrt{x}A^\ast) &= \text{Im}\left[\sqrt{x}A \,(u^\ast)^2\right].
       \end{split}
       \end{equation}
       For general values of $u \in \mathds{C}$,
       the limiting cases in Fig.~\ref{fig:cuts-4} correspond to
       \begin{itemize}
       \item $C^2 > 0$ (top left): $A=\lambda iu$ and $\lambda^2 < 1$,
       \item $C^2 < 0$ (bottom right): either $A=\lambda iu$ and $\lambda^2 > 1$, or $A=\lambda u$,
       \end{itemize}
       where $\lambda \in \mathds{R}$.
       The location of $u$ then divides
       the complex plane into four quadrants delimited by the lines $\lambda iu$ and $\lambda u$,
       where the cuts generated by the points $A$ inside these quadrants take intermediate shapes like in Fig.~\ref{fig:cuts-4}.
       The situation for $iu$ in the upper right quadrant is illustrated in Fig.~\ref{fig:cuts-5}: depending on whether $\arg{(A)} < \arg{(iu)}$ or $\arg{(A)} > \arg{(iu)}$,
       the cuts twirl in one or the other direction.

       Writing $iu=g+ih$ with $g,h\in\mathds{R}$,  $C^2 = -1 - u^2/A^2$ entails
       \begin{equation}
       \begin{split}
           d^2-c^2 &= 1 + \frac{(a^2-b^2)(h^2-g^2) - 4abgh}{|A|^4}\,, \\
          cd &= \frac{ab\,(h^2-g^2) + gh\,(a^2-b^2)}{|A|^4}\,, \\
          d^2-c^2-2cdr &= 1 - \frac{h^2-g^2+2ghr}{|A|^2}\,, \\
          (c+dr)(d-cr) &= r-\frac{gh}{b^2}\,,
       \end{split}
       \end{equation}
       so that Eq.~\eqref{curve-crosses-zero} for the zero crossing results in
       \begin{equation}\label{curve-crosses-zero-2}
       \begin{split}
          |\sqrt{x}_\pm| &=  \sqrt{|A|^2 - (h^2-g^2+2ghr)}\,, \\
          |\omega| &= \left| a - \frac{gh}{b}\right| / |\sqrt{x}_\pm|\,.
       \end{split}
       \end{equation}
       As long as $|A| < \sqrt{h^2-g^2+2ghr}$, the cuts do not cross the real axis.

       Also relevant is the condition for the cuts to cross the real axis if $\omega$ is restricted to the interval $\omega \in [-1,1]$.
       In that case Eq.~\eqref{curve-crosses-zero-2} must be satisfied for $|\omega| < 1$, which implies $b^2 > h^2$; or in other words,
       the cuts do not cross the real axis as long as $|\text{Im}\,A| < |\text{Im}\,iu|$.

       Another question concerns the general singularity locations in Eq.~\eqref{kinematics-E2},
       $q_{1,2}^2 = -m^2 u^2$ for the propagators and  $q^2 = -m^2 u^2$ for the BS amplitude, with $u \in \mathds{C}$.
       This is relevant for the propagators entering in $G_0(q,P)$, because in general  they may not be known in the whole complex plane,
       and for the BS amplitude $\Gamma(q,P)$ which may dynamically generate singularities in the course of the BSE solution.
       Both cases translate to Eq.~\eqref{cut-function-2} and Fig.~\ref{fig:cuts-5} for $iu_k$ ($k=1,2,3, \dots$) in the upper right quadrant.
       Any such singularity generates a branch cut in the complex $\sqrt{x}$ plane, where the region $R_\epsilon$ in Fig.~\ref{fig:cuts-7}
        arises from the intersection of all cuts for a given value  $A \propto \sqrt{t}$.
        For example, if the amplitude generates only real singularities (then the respective $iu_k$ are imaginary)
        but the propagators produce complex singularities (so that $iu_k  \in \mathds{C}$),
        then as long as
        \begin{equation}\label{u-condition}
           \arg(A)  < \min\left\{\arg(iu_k)\right\}
        \end{equation}
        a contour deformation is always possible because there is always a path in $\sqrt{x} \in \mR_\epsilon$ that connects the origin with infinity.
        Vice versa, for $\arg(A)  > \min\left\{\arg(iu_k)\right\}$ like in the right panel of Fig.~\ref{fig:cuts-5},
        the different cuts may intersect such that no such path can be found.
        In the extreme case  where some singularity $iu$ moves to the positive real axis,
        the equations can only be solved for $\sqrt{t}\in \mathds{R}_+$, whereas
        in the opposite extreme case where all singularities $iu_k$ are confined to the imaginary axis like in Fig.~\ref{fig:cuts-7},
        a contour deformation is always possible and the equations can be solved for any $\sqrt{t} \in \mathds{C}$.

  \newpage

\bibliographystyle{apsrev4-1-mod}
\bibliography{bib-lf}

\end{document}